\documentclass{article}

\usepackage{jheppub}

\usepackage{bbm}
\usepackage{booktabs}
\usepackage{braket}
\usepackage{comment}
\usepackage{dcolumn}
\usepackage{dsfont}
\usepackage{graphicx}
\usepackage[utf8]{inputenc}
\usepackage{multirow}
\usepackage{rotating}
\usepackage{siunitx}
\usepackage{slashed}
\usepackage{subcaption}
\usepackage[normalem]{ulem}
\usepackage{xcolor}
\usepackage{xspace}
\usepackage{hyperref}
\usepackage[nameinlink,capitalize]{cleveref}
\usepackage{tabularx}

\newcommand{\eq}[1]{\begin{equation} #1 \end{equation}}
\newcommand{\eqa}[1]{\begin{eqnarray} #1 \end{eqnarray}}
\newcolumntype{C}{>{\centering\arraybackslash}X}

\DeclareMathOperator{\Tr}{Tr}

\renewcommand{\L}{{\cal L}}
\renewcommand{\O}{{\cal O}}

\begin{document}

\setcounter{tocdepth}{2}

\title{Inclusive Charmless Non-Leptonic $\boldsymbol{B}$ Decays at NLO\\ within and beyond the Standard Model}

\author[a,b]{Stefan Meiser,}
\author[a,b]{Pol Morell,}
\author[a,b]{Angel Picazo,}
\author[a,b]{Javier Virto}

\emailAdd{smeiser@icc.ub.edu}
\emailAdd{pmorell@icc.ub.edu}
\emailAdd{apicazar7@icc.ub.edu}
\emailAdd{jvirto@icc.ub.edu}
\affiliation[a]{Departament de Física Quàntica i Astrofísica, Universitat de Barcelona,\\
Martí i Franquès 1, E08028 Barcelona, Catalonia}
\affiliation[b]{
Institut de Ciències del Cosmos (ICCUB), Universitat de Barcelona,\\
Martí i Franquès 1, E08028 Barcelona, Catalonia}

\abstract{%
We calculate the full set of BSM contributions to the inclusive non-leptonic $B$-meson lifetime from charmless final states within the framework of the Weak Effective Theory up to next-to-leading order in QCD and to leading power in the heavy-quark expansion. We do so by computing cut diagrams and the corresponding phase space integrals. This involves calculating current-current, penguin-penguin, penguin-dipole and dipole-dipole diagrams with two-, three- and four-particle cuts. We describe the technical difficulties of the computation.  
We then discuss how our results can be used to constrain new physics scenarios related to the anomalies observed in semi-leptonic and charmless non-leptonic decays.
}

\vspace*{-3\baselineskip}

\maketitle

\allowdisplaybreaks

\section{Introduction}
\label{sec:intro}

Lifetimes of weakly-decaying $B$-mesons, e.g. $B^\pm, B^0, B^0_s, B^\pm_c$, are known experimentally to a high precision, currently surpassing the per-mille level in several cases (see, for instance, Ref.~\cite{ParticleDataGroup:2024cfk}).
Theoretically, the lifetimes of $B_q$-mesons with a light spectator quark $q$ can be systematically calculated within the Heavy Quark Expansion (HQE) (see Ref.~\cite{Shifman:1986mx} for the seminal paper on the topic, and Ref.~\cite{Lenz:2014jha} for a modern review.) 
Within this framework, the $B$-meson decay width can be expressed as the sum of the free $b$-quark decay width $\Gamma(b)$ plus further spectator-dependent power corrections $\delta \Gamma(B_q)$,
\eqa{
\Gamma(B_q) = \Gamma(b) + \delta\Gamma(B_q) \, .
}
While power corrections in $\delta \Gamma(B_q)$ are suppressed by at least two powers of $\Lambda_\text{QCD}/m_b$, suggesting that the free-quark decay should approximate the meson lifetime well, at low orders in QCD $\Gamma(b)$ is plagued by large scale uncertainties~\cite{Albrecht:2024oyn}. Thus, considerable theoretical efforts have been made in taming the scale uncertainty by computing QCD corrections to the free-quark decay within the Standard Model~(SM) up to next-to-next-to-leading order (NNLO) precision in non-leptonic channels~\cite{Bagan:1994zd,Voloshin:1994sn,Bagan:1995yf,Lenz:1997aa,Lenz:1998qp,Greub:2000an,Greub:2000sy,Czarnecki:2005vr,Krinner:2013cja,Egner:2024azu} and next-to-next-to-next-to-leading order ($\text{N}^3\text{LO}$) in semi-leptonic ones~\cite{Czarnecki:1997hc,Czarnecki:1998kt,vanRitbergen:1999gs,Melnikov:2008qs,Pak:2008cp,Pak:2008qt,Dowling:2008ap,Bonciani:2008wf,Biswas:2009rb,Brucherseifer:2013cu,Egner:2023kxw,Fael:2020tow,Czakon:2021ybq,Fael:2023tcv,Broggio:2026edk}.
While the resulting theoretical predictions agree well with data in general, theoretical uncertainties are still much larger than the experimental ones.

As a complementary approach, one can consider lifetime ratios, where the free-quark decay rate drops out (and so does the associated scale uncertainty) and only power corrections contribute 
\eqa{
\frac{\tau(B_q)}{\tau(B_{q^\prime})} = 1 + \left[\delta \Gamma(B_{q^\prime}) - \delta \Gamma(B_q) \right] \cdot \tau(B_q) \, .
}
SM predictions of these ratios successfully reproduce experimental data~\cite{Bigi:1992ne,Neubert:1996we,Lenz:2014jha,Kirk:2017juj,Moretti:2026waq}, and thus can be used to constrain physics Beyond the SM (BSM)~\cite{Lenz:2022pgw,Black:2024bus,Lang:2025ios}.
In the case of $b \to c \bar{u}d(s)$ transitions, where some tensions between SM predictions and experimental data are observed \cite{Bordone:2020gao,Cai:2021mlt,Meiser:2024zea,Atkinson:2024hqp}, the lifetime ratios have been computed for the full set of operators in the so-called Weak Effective Theory (WET) \cite{Lenz:2022pgw,Lang:2025ios} and employed as complementary constraints for ruling out different NP scenarios addressing these tensions \cite{Lenz:2022pgw,Atkinson:2024hqp,Araz:2026zlu}.
However, as pointed out in Ref.~\cite{Lenz:2022pgw}, these lifetime ratios leave some of the directions in the space of $b \to c\bar{u}d(s)$ Wilson coefficients (WC) unconstrained. 
It is thus convenient in this context to use the full lifetimes (not the ratios), which provide absolute bounds in the space of~WCs. This has been demonstrated in Ref.~\cite{Meiser:2024zea}, where the $b \to c\bar{u}d(s)$ lifetimes were computed at LO in QCD and at leading power in the HQE in the full WET, leading to constraints in WC space defining a compact region with no flat directions in the relevant sector.

Motivated by the results of Ref.~\cite{Meiser:2024zea}, we set out to improve this LO calculation by computing the contributions to the free $b$-quark decay from all possible non-leptonic channels up to next-to-leading order in QCD. This work represents a first step in this endeavor: the complete NLO calculation of the parton-level decay for charmless non-leptonic final states. Contributions from final states with charm are technically more challenging and are left for future work.

The calculation of non-leptonic $B$-meson observables within the most general effective Lagrangian (beyond the SM operator set) is also part of a generalized endeavor to systematize and automate phenomenological analyses of experimental data through the formalism of Effective Field Theory~(EFT)~\cite{Aebischer:2023nnv}. 
For observables at energies lower than the EW scale, such as $B$-decay observables, the relevant EFT is known as the Low-Energy EFT (LEFT)~\cite{Jenkins:2017jig} or the Weak Effective Theory (WET)~\cite{Aebischer:2017gaw}. 
This framework allows for an efficient connection between low-energy and high-energy observables, extending beyond the electroweak scale onto the Standard Model EFT (SMEFT)~\cite{Grzadkowski:2010es,Isidori:2023pyp}. Indeed, much effort is currently being devoted into setting up the EFT pipeline at the level of next-to-leading logarithms~(NLL), by computing the complete set of two-loop Renormalization Group Equations (RGEs) in the LEFT~\cite{Morell:2024aml,Naterop:2024cfx,Aebischer:2025hsx,Naterop:2025lzc,Naterop:2025cwg} and the SMEFT~\cite{Banik:2025wpi,Born:2026xkr}, and the full one-loop matching conditions to the SMEFT~\cite{Dekens:2019ept}. 
This pipeline is also being implemented in automated tools, such as \texttt{DsixTools}~\cite{Celis:2017hod,Fuentes-Martin:2020zaz} and \texttt{wilson}~\cite{Aebischer:2018bkb}.
The NLL EFT set-up needs then to be matched by the calculation of observables to NLO.

The structure of this article is as follows. In~Section~\ref{sec:WET} we introduce the definitions for the relevant sectors of the WET that are needed in the calculation. In~Section~\ref{sec:Calculation}, we discuss the HQE, the difficulties that arise in our calculation when treating traces with $\gamma_5$ within the renormalization scheme defined by naive dimensional regularization (NDR), and the computation of the loop and phase space integrals.
We then present our results in~Section~\ref{sec:results}, and perform a numerical analysis in~Section~\ref{sec:applications} to assess their potential phenomenological impact. We conclude in~Section~\ref{sec:conclusions}.
Some of the more technical material is relegated to a few appendices.

\section{The Weak Effective Lagrangian}
\label{sec:WET}
Our computation focuses on the contributions to the non-leptonic $B$-meson lifetime due to light final states (containing only $u,d$ and $s$ quarks, and gluons). This includes four distinct cases, namely:
\begin{enumerate}
    \item One quark-antiquark pair and an additional quark with different flavor in the final state ($b\to{}d\bar{u}u$, $b\to{}d\bar{s}s$, $b\to{}s\bar{u}u$, $b\to{}s\bar{d}d$).
    \item Two identical quarks and an antiquark with a different flavor in the final state ($b\to{}d\bar{s}d$, $b\to{}s\bar{d}s$).
    \item A set of quarks with three identical flavors in the final state ($b\to{}d\bar{d}d$, $b\to{}s\bar{s}s$).
    \item A quark and a gluon in the final state ($b\to{}dg$, $b\to{}sg$).
\end{enumerate}
For definiteness, we shall only consider explicitly the cases $b \to d \bar u u$, $b \to d g$, $b \to d \bar s d$  and $b \to d \bar d d$, with the other cases obtained trivially by a change of flavor.
Let us also note that the calculation will also necessarily include four-particle contributions ($b\to d\bar q_2 q_3 g$), as explained below.

\subsection{EFT sector $b \to d \bar u u$, $b \to d \bar d d$ and $b \to d g$}

Cases 1, 3 and 4 are all described by the same set of 82 effective operators in the EFT. The corresponding effective Lagrangian can be written as
\eqa{
\L^{dbX} = \L^{dbuu} + \L^{dbss} + \L^{dbcc} + \L^{dbdd} + \L^{dbbb} + \L^{dbG}\ ,
}
where the terms $\L^{dbqq}$ ($q=u,s,c,d,b$) contain all four-quark operators and $\L^{dbG}$ contains the two QCD dipole operators. Throughout this work, and for the reasons discussed in~Section~\ref{sec:gamma5}, we will use the so-called {\it Bern operator basis}~\cite{Aebischer:2017gaw}.
The relevant operators are the ones classified as Class V in~Ref.~\cite{Aebischer:2017gaw}, excluding the semi-leptonic and QED dipole operators. For reference, we explicitly quote them here:
\eqa{
\label{eq:Lagrangian_dbqq}
\L^{dbqq} = \frac{4G_F}{\sqrt{2}} \sum_{i = 1}^{10} \left( \mathcal{C}^{dbqq}_i \mathcal{O}^{dbqq}_i + \mathcal{C}^{dbqq}_{i^\prime} \mathcal{O}^{dbqq}_{i^\prime} \right) \, ,
}
where, whenever $q=u,s,c$, 
\begin{align}
\mathcal{O}_1^{dbqq} &= \left[ \bar{d} P_R \gamma_\mu b \right] \left[ \bar{q} \gamma^\mu q \right] \, , & 
\mathcal{O}_2^{dbqq} &= \left[ \bar{d} P_R \gamma_\mu T^A b \right] \left[ \bar{q} \gamma^\mu T^A q \right]  \, , \notag 
\\
\mathcal{O}_3^{dbqq} &= \left[ \bar{d} P_R \gamma_{\mu \nu \rho} b \right] \left[ \bar{q} \gamma^{\mu \nu \rho} q \right]\, , &
\mathcal{O}_4^{dbqq} &= \left[ \bar{d} P_R \gamma_{\mu \nu \rho} T^A b \right] \left[ \bar{q} \gamma^{\mu \nu \rho} T^A q \right] \, , \notag 
\\
\mathcal{O}_5^{dbqq} &= \left[ \bar{d} P_R b \right] \left[ \bar{q} q \right]\, , & 
\mathcal{O}_6^{dbqq} &= \left[ \bar{d} P_R T^A b \right] \left[ \bar{q} T^A q \right] \, , 
\label{eq:Bern} \\
\mathcal{O}_7^{dbqq} &= \left[ \bar{d} P_R \sigma_{\mu \nu} b \right] \left[ \bar{q} \sigma^{\mu \nu} q \right]\, , &
\mathcal{O}_8^{dbqq} &= \left[ \bar{d} P_R \sigma_{\mu \nu} T^A b \right] \left[ \bar{q} \sigma^{\mu \nu} T^A q \right] \, , \notag 
\\
\mathcal{O}_9^{dbqq} &= \left[ \bar{d} P_R \gamma_{\mu \nu \rho \sigma} b \right] \left[ \bar{q} \gamma^{\mu \nu \rho \sigma} q \right]\, , &
\mathcal{O}_{10}^{dbqq} &= \left[ \bar{d} P_R \gamma_{\mu \nu \rho \sigma} T^A b \right] \left[ \bar{q} \gamma^{\mu \nu \rho \sigma} T^A q \right] \, . \notag 
\end{align}
We use the same conventions as in~Ref.~\cite{Aebischer:2017gaw}, in particular $\sigma_{\mu \nu} \equiv \frac{i}2 [\gamma_\mu, \gamma_\nu]$, $P_{L,R} \equiv \frac12(1 \mp \gamma_5)$, and $\gamma_{\mu_1 \dots \mu_n} \equiv \gamma_{\mu_1} \cdots \gamma_{\mu_n}$.
The other ten operators, denoted by primed indices, are obtained by replacing $P_L \leftrightarrow P_R$ in the definitions above. For $q = d,b$ the color-octet operators are Fierz-redundant and hence not included in the basis; therefore $\mathcal{O}_i^{dbdd}$ and $\mathcal{O}_i^{dbbb}$ exist only for odd $i$. Furthermore, for $q = d$, the operators $\O^{dbdd}_{5, 7, 9}$ are defined with a left-handed projector ($P_L$) instead of the right-handed one ($P_R$), which greatly simplifies the mixing pattern.
The Lagrangian for the QCD dipole operators reads
\eqa{
\label{eq:Lagrangian_dbG}
\mathcal{L}^{dbG} = \frac{4 G_F}{\sqrt{2}} \left(\mathcal{C}_{8g}^d \mathcal{O}_{8g}^d + \mathcal{C}_{8^\prime g}^d \mathcal{O}_{8^\prime g}^d \right) \, ,
}
with
\eqa{
\label{eq:OP_dipoles}
\mathcal{O}^d_{8^{(\prime)}g} = \frac{m_b}{g_s} \left[\bar{d} \sigma^{\mu \nu} P_{R(L)} T^A b \right] G^A_{\mu \nu} \, .
}
Our convention for the covariant derivative is the same as in~Ref.~\cite{Aebischer:2017gaw}. 

As opposed to the Wilson coefficients of four-quark operators, for which we take an agnostic approach with regard to their magnitude, we shall henceforth consider all dipole-operator coefficients, in this case ${C}_{8^{(\prime)}g}^{d}$, as implicitly suppressed by a factor of at least $1/16\pi^2$, thus contributing at the same level as one-loop diagrams in the loop expansion. 
We choose to do so because the SMEFT matches at tree-level to these coefficients only through the SMEFT QCD dipoles, $Q_{dG}$, and the matching conditions from all other operators (including the SM matching) arise only starting at one-loop order~\cite{Dekens:2019ept}. 
Furthermore, it is a well-known fact that typical UV completions of the SM do not match onto dipole operators in the SMEFT at tree level~\cite{Arzt:1994gp,deBlas:2017xtg}.

For later convenience, we introduce the compact notation $\mathcal{C}^{dbX}_i$, with index $i=1,\, \dots \, ,41$: 
\begin{equation}
    \big\{\mathcal{C}^{dbX}_i\big\} \equiv \big\{\mathcal{C}_{1-10}^{dbuu},\, \mathcal{C}_{1-10}^{dbss}, \, \mathcal{C}_{1-10}^{dbcc}, \mathcal{C}_{1-9}^{dbdd}, \, \mathcal{C}_{1-9}^{dbbb}, \, \mathcal{C}^d_{8g}\big\} \, .
\end{equation}
Note that the normalization of the Lagrangian in~Eqs.~(\ref{eq:Lagrangian_dbqq}) and~(\ref{eq:Lagrangian_dbG}) only contains the Fermi constant and numerical factors, but not the CKM matrix elements, which are instead absorbed into the definition of the Wilson coefficients.

\subsection{EFT sector for $b\to d \bar s d$}
The sector of the effective Lagrangian relevant for the process $b \to d \bar{s} d$ contains 10 four-quark operators:
\eqa{
\label{eq:Lagrangian_dbds}
\mathcal{L}^{dbds} = \frac{4 G_F}{\sqrt{2}} \sum_{\substack{i=1 \\ i \text{ odd}}}^{10} \left(\mathcal{C}_i^{dbds} \mathcal{O}_i^{dbds} + \mathcal{C}_{i^\prime}^{dbds} \mathcal{O}_{i^\prime}^{dbds} \right) \, .
}
In the Bern basis this sector consists of the operators classified as Class IV in~Ref.~\cite{Aebischer:2017gaw}:
\begin{align}
\mathcal{O}_1^{dbds} &= \left[ \bar{d} \gamma_\mu P_L b \right] \left[ \bar{d} \gamma^\mu s \right] \, , & 
\mathcal{O}_{1'}^{dbds} &= \left[ \bar{d} \gamma_\mu P_R b \right] \left[ \bar{d} \gamma^\mu s \right]  \, , \notag 
\\
\mathcal{O}_3^{dbds} &= \left[ \bar{d} \gamma_{\mu \nu \rho} P_L b \right] \left[ \bar{d} \gamma^{\mu \nu \rho} s \right]\, , &
\mathcal{O}_{3'}^{dbds} &= \left[ \bar{d} \gamma_{\mu \nu \rho} P_R b \right] \left[ \bar{d} \gamma^{\mu \nu \rho} s \right] \, , \notag 
\\
\mathcal{O}_5^{dbds} &= \left[ \bar{d} P_L b \right] \left[ \bar{d} s \right]\, , & 
\mathcal{O}_{5'}^{dbds} &= \left[ \bar{d} P_R b \right] \left[ \bar{d} s \right] \, , 
\label{eq:Berndbds} \\
\mathcal{O}_7^{dbds} &= \left[ \bar{d} \sigma_{\mu \nu} P_L b \right] \left[ \bar{d} \sigma^{\mu \nu} s \right]\, , &
\mathcal{O}_{7'}^{dbds} &= \left[ \bar{d} \sigma_{\mu \nu} P_R b \right] \left[ \bar{d} \sigma^{\mu \nu} s \right] \, , \notag 
\\
\mathcal{O}_9^{dbds} &= \left[ \bar{d} \gamma_{\mu \nu \rho \sigma} P_L b \right] \left[ \bar{d} \gamma^{\mu \nu \rho \sigma} s \right]\, , &
\mathcal{O}_{9'}^{dbds} &= \left[ \bar{d} \gamma_{\mu \nu \rho \sigma} P_R b \right] \left[ \bar{d} \gamma^{\mu \nu \rho \sigma} s \right] \, . \notag 
\end{align}
Again, there are only color-singlet operators since the color-octet operators are Fierz-redundant.

\section{Outline and Details of the Calculation}
\label{sec:Calculation}

In the heavy-quark limit the decay rate of the $B$-meson is equal to the parton-level $b$-quark decay rate~\cite{Shifman:1986mx,Lenz:2014jha},
\eqa{
\Gamma(B) = \Gamma(b) + \mathcal{O}(1/m_b^2) \, .
}
This relation receives power corrections that are calculable within the heavy-quark expansion (HQE).
The partonic contribution to the ``non-leptonic'' $B$-meson lifetime $\Gamma^{\text{NL}}(B)$ to leading order in QED is given by the decay rate of the $b$-quark to quarks and gluons, $\Gamma^{\text{NL}}(b)$.
Given that such final states cannot be traced back to specific hadronic final states, the connection back to the $B$-meson decay rate can only be achieved at the inclusive level.
We evaluate it using the optical theorem
\eqa{
\Gamma^{\text{NL}}(b) = \frac{1}{2m_b} \text{Im}\, \big\langle b\big| \, i \int d^4x \, T \{\mathcal{L}(x) \mathcal{L}(0)\}\big|b\big\rangle \, , 
\label{eq:optTh}
}
with the relevant sectors of the effective Lagrangian included in $\mathcal{L}$. 
We focus exclusively on the contributions due to \textit{massless} intermediate states, i.e. $u,d,s$ quarks, and gluons. 

In order to evaluate~Eq.~(\ref{eq:optTh}), we need to compute the imaginary parts of one-, two- and three-loop diagrams, as described in Sec.~\ref{sec:Diagrams}.
We do so by applying Cutkosky rules (see Appendix~\ref{app:Cutkosky}) to the loop integrals, which transform loop integrals with a cut into $n$-particle phase-space integrals. 
The NLO calculation requires two-particle $q_1g$ cuts, three-particle $q_1 \bar{q}_2 q_3$ cuts, and four-particle $q_1\bar{q}_2q_3 g$ cuts.
With this in mind, the contribution from an $n$-particle cut to the decay width can be written as
\eqa{
    \Gamma^{(n)}(b) &=& 8 G_F^2 \frac{1}{2N_c} \frac{1}{2m_b} \, \mathcal{C}_i^\dagger \mathcal{C}_j \int d\Pi_n \, \mathcal{K}^{(n)}_{ij} \, ,
}
where $d\Pi_n$ is the $n$-particle phase-space measure. The integration kernels $\mathcal{K}^{(n)}_{ij}$ arise from the calculation of the set of $n$-particle cut diagrams, as discussed in what follows.

%
\begin{figure}[t]
  \centering
  \begin{subfigure}{0.45\textwidth}
    \centering
    \includegraphics[width=\textwidth]{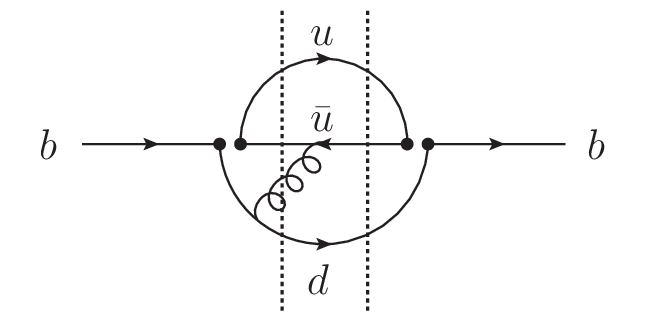}
    \label{fig:CC}
  \end{subfigure}
  \hfill 
  \begin{subfigure}{0.45\textwidth}
    \centering
    \raisebox{1.25cm}{
\includegraphics[width=\textwidth]{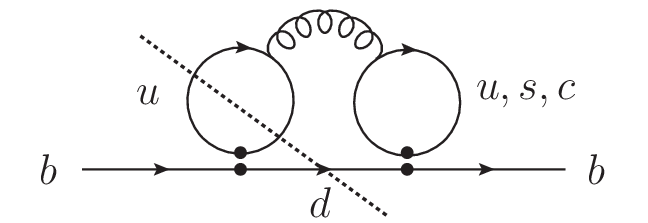}}
    \label{fig:PenguinPenguinCC}
  \end{subfigure}

  \vspace{0.1cm} 

  \begin{subfigure}{0.45\textwidth}
    \centering
    \includegraphics[width=\textwidth]{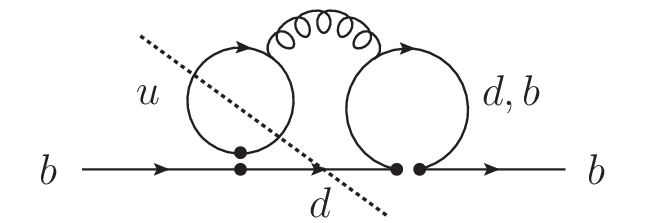}
    \label{fig:PenguinDipole}
  \end{subfigure}
  \hfill
  \begin{subfigure}{0.45\textwidth}
    \centering
    \includegraphics[width=\textwidth]{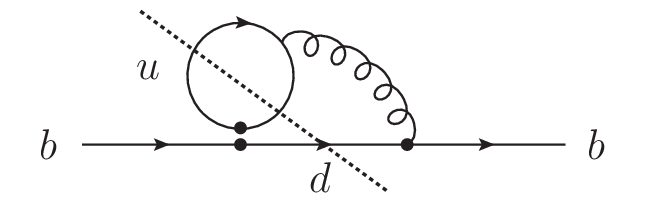}
    \label{fig:DipoleDipole}
  \end{subfigure}

  \caption{Topologies that contribute to the $b \to d \bar{u} u$ decay rate at NLO. The dotted lines represent the $u\bar u d(g)$ three- and four-particle cuts. Top row: $CC$ straight and $PP$ closed-closed topologies. Bottom row: $PP$ closed-open and $PD$ closed topologies. The hermitian conjugated diagrams are not shown.}
  \label{fig:Topologies1}
\end{figure}

\subsection{Set of Diagrams}
\label{sec:Diagrams}

\begin{figure}[t]
  \centering
  \begin{subfigure}{0.45\textwidth}
    \centering
    \includegraphics[width=\textwidth]{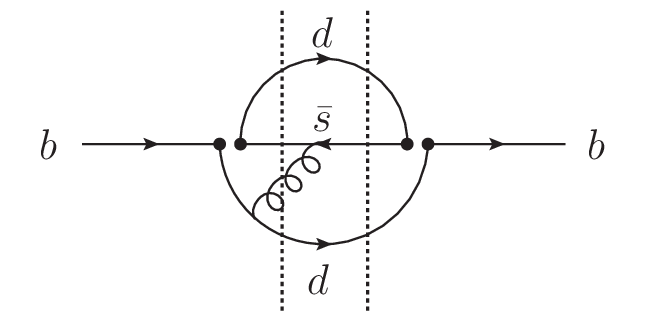}
    \label{fig:CCdbds}
  \end{subfigure}
  \hfill
  \begin{subfigure}{0.45\textwidth}
    \centering
    \includegraphics[width=\textwidth]{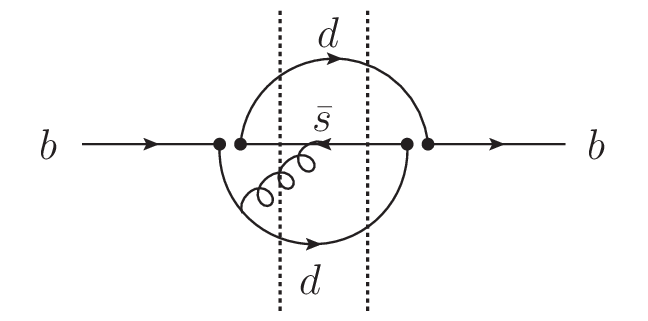}
    \label{fig:Fierzeddbds}
  \end{subfigure}
  \caption{Topologies that contribute to the $b \to d \bar{s} d$ decay rate at NLO. The dotted lines represent the $d\bar{s}d$ three- and four-particle cuts. There are only the straight and crossed $CC$ insertions.}
  \label{fig:Topologies2}
\end{figure}
\begin{figure}[t]
  \centering
  
  \begin{subfigure}{0.32\textwidth}
    \centering
    \includegraphics[width=\textwidth]{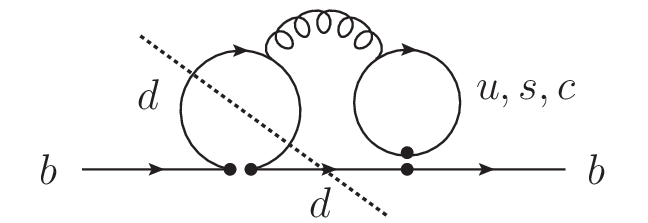}
    \label{fig:PPOC}
  \end{subfigure}
  \hfill
  \begin{subfigure}{0.32\textwidth}
    \centering
    \includegraphics[width=\textwidth]{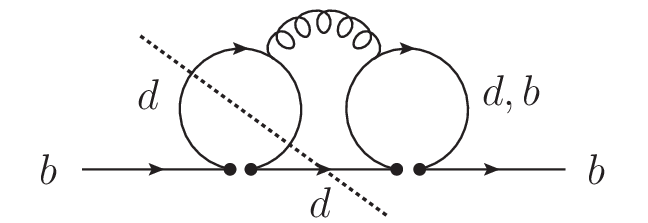}
    \label{fig:PPOO}
  \end{subfigure}
  \hfill
  \begin{subfigure}{0.32\textwidth}
    \centering
    \includegraphics[width=\textwidth]{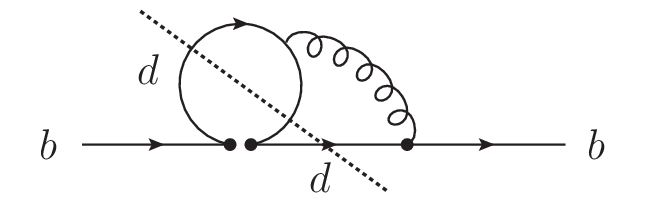}
    \label{fig:DPO}
  \end{subfigure}
  \caption{$PP$ and $PD$ topologies contributing to the $b \to d \bar{d} d$ decay rate at NLO. The dotted lines represent the $d\bar{d}d$ cuts. $CC$ diagrams are the same as in $b \to d \bar{s}d$ (Fig.~\ref{fig:Topologies2}) with the replacement $\bar s \to \bar d$.}
  \label{fig:Topologies3}
\end{figure}
\begin{figure}[t]
  \centering
  \begin{subfigure}{0.32\textwidth}
    \centering
    \includegraphics[width=\textwidth]{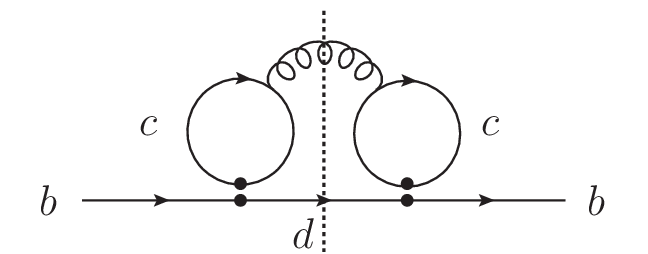}
    \label{fig:PPCC2pt}
  \end{subfigure}
  \hfill 
  \begin{subfigure}{0.32\textwidth}
    \centering
    \includegraphics[width=\textwidth]{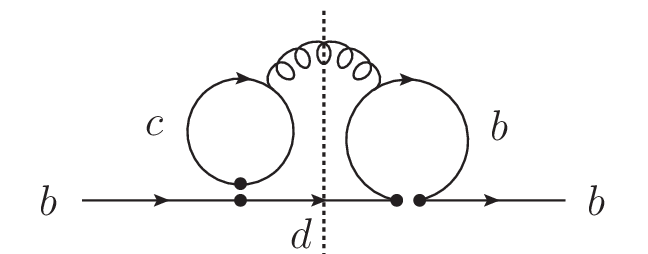}
    \label{fig:PPCO2pt}
  \end{subfigure}
  \hfill
  \begin{subfigure}{0.32\textwidth}
    \centering
    \includegraphics[width=\textwidth]{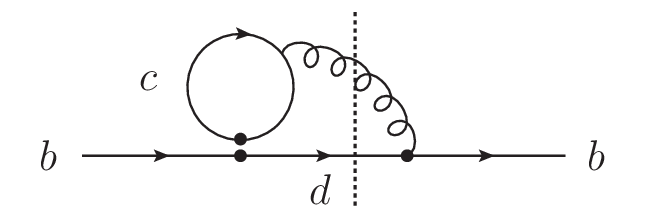}
    \label{fig:PD2pt}
  \end{subfigure}
  \begin{subfigure}{0.32\textwidth}
    \centering
    \includegraphics[width=\textwidth]{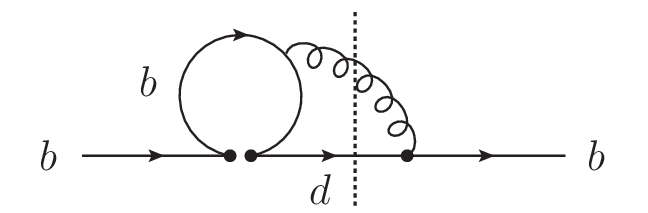}
    \label{fig:PDO2pt}
  \end{subfigure}
  \begin{subfigure}{0.32\textwidth}
    \centering
    \includegraphics[width=\textwidth]{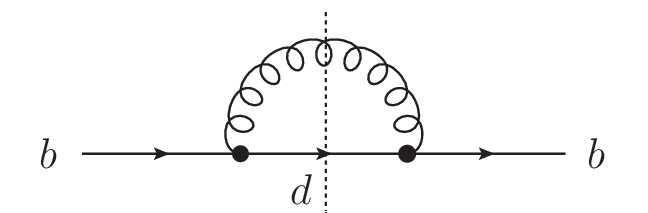}
    \label{fig:DD}
  \end{subfigure}
  \caption{Topologies that contribute to the $b \to d g$ decay rate at NLO. The dotted lines represent the $dg$ two-particle cuts. Only massive quarks in the penguin loops lead to non-vanishing contributions.}
  \label{fig:Topologies4}
\end{figure}

The calculation involves insertions of the relevant effective operators into current-current~($CC$), penguin-penguin~($PP$), penguin-dipole~($PD$), and dipole-dipole~($DD$) topologies, depending on the case:
\paragraph{\textbf{Case 1}} $(b\rightarrow{d \bar{u} u})$: As shown in~Fig.~\ref{fig:Topologies1}, the $b \to d\bar{u}u$ rate receives contributions from $CC$, $PP$, and $DP$ topologies. 
For the $CC$ topologies, only the straight insertion, where there are two Dirac traces, arises.
These diagrams contain three-particle ($d \bar u u$) and four-particle ($d \bar{u} u g$) cuts.
The $PP$ and $DP$ topologies only contribute via three-particle cuts to the $b \to d\bar{u}u$ rate.
\paragraph{\textbf{Case 2}} $(b\rightarrow{d \bar{s} d})$: For $b \to d \bar s d$, there are only contributions from straight and crossed $CC$ topologies as illustrated in~Fig.~\ref{fig:Topologies2}, again featuring three- and four-particle cuts. 
Diagrammatically, the straight insertion is the same as in Case 1, but with the replacements $u \rightarrow d$ and $\bar{u} \rightarrow \bar{s}$, while the crossed insertion has only one trace.  
%
%
%
\paragraph{\textbf{Case 3}} $(b\rightarrow{d \bar{d} d})$: There are contributions from crossed and straight $CC$, open and closed $PP$, and open $PD$ topologies.
The $PP$ and $PD$ insertions are shown in~Fig.~\ref{fig:Topologies3}, while the $CC$ insertions are the same as in~Fig.~\ref{fig:Topologies2} with the substitution $\bar s \to \bar d$.
%
%
\paragraph{\textbf{Case 4}} $(b\rightarrow{d g})$: As illustrated in~Fig.~\ref{fig:Topologies4}, one needs to compute the same $PP$ and $DP$ diagrams as in Case 1 and Case 3, but only performing the two-particle cuts.
Additionally, there is a contribution from a $DD$ topology.

%
\subsection{Treatment of $\gamma_5$}
\label{sec:gamma5}

\begin{figure}[t]
  \centering
  \begin{subfigure}{0.45\textwidth}
    \centering
    \includegraphics[width=\textwidth]{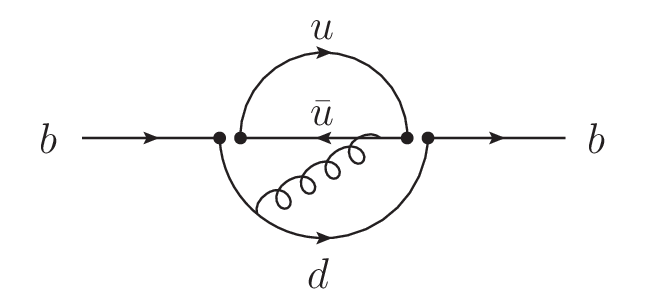}
    \label{fig:CCwoCuts}
  \end{subfigure}
  \hfill
  \begin{subfigure}{0.45\textwidth}
    \centering
    \includegraphics[width=\textwidth]{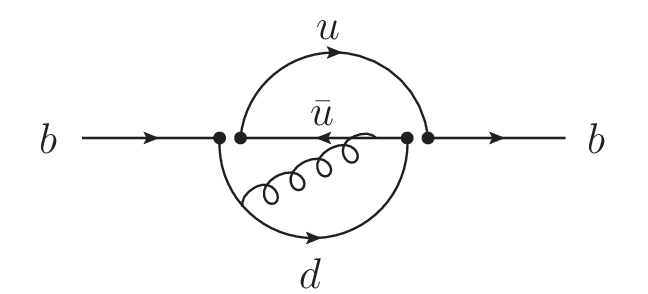}
    \label{fig:FierzedwoCuts}
  \end{subfigure}

  \caption{Effect of the Fierz transformation acting on the operator inserted at the right vertex using the example of a NLO straight $CC$ diagram that contributes to the $b \to d \bar{u} u$ decay rate. In the Bern basis the left diagram is calculable without problems while in more conventional bases it would lead to ambiguous results due to traces with $\gamma_5$. Hence, one would have to Fierz transform one of the operators, yielding the crossed $CC$ diagram on the right with just one Dirac trace.}
  \label{fig:FierzEffect}
\end{figure}

As can be seen in the first diagram in~Fig.~\ref{fig:FierzEffect}, a straight $CC$ insertion leads to a contribution with two Dirac traces. This is problematic when computing such diagrams in NDR, as originally pointed out in~Refs.~\cite{Bagan:1994zd,Bagan:1995yf,Egner:2024azu,Mannel:2025fvj,Moretti:2026waq}, since in this scheme traces of the form $\Tr(\gamma^{\mu_1}\gamma^{\mu_2}\gamma^{\mu_3}\gamma^{\mu_4}\cdots\gamma_5)$ are ambiguous.
We can use the fact that the inclusive decay width must be parity-even, while a single Dirac trace with an odd number of $\gamma_5$ is not.
Then, any term contributing to the decay rate with a single such trace can be instantly discarded~\cite{Huber:2014nna}. However, a product of two traces containing a $\gamma_5$ each will yield a parity-even contribution, which we do have to include. There are mainly two ways of dealing with this problem within NDR.

\begin{enumerate}

\item The first solution is to Fierz-transform one of the inserted operators such that the diagram produces only one Dirac trace, as can be seen in the second diagram of~Fig.~\ref{fig:FierzEffect}. 
We can then discard all terms featuring a trace with $\gamma_5$. 
This, however, leads to a different contribution, yielding the decay width in terms of the Wilson coefficients of the original basis on one side, but in terms of the Wilson coefficients of the Fierz-transformed operators on the other side. In previous works addressing the SM predictions for these and similar decays~\mbox{\cite{Buras:1989xd,Bagan:1994zd,Bagan:1995yf,Egner:2024azu,Mannel:2025fvj,Moretti:2026waq}} this problem was solved through an appropriate choice of evanescent operators that ensures that Fierz symmetry is preserved at NLO. This allows one to freely substitute the Wilson coefficients of the Fierz-transformed operators with their untransformed counterparts using a tree-level change of basis (i.e. a simple rotation of the basis). 
In our case we are interested in the full set of BSM contributions, and we need to be more careful when relating the Fierz-transformed Wilson coefficients to the original ones, since the standard choice of evanescent operators for the full BSM basis does not preserve Fierz symmetry at NLO. If one is then to treat this approach properly, one needs to address this method within the framework of a change of basis to NLO on the Fierz-transformed WCs. We will explore this procedure in depth in a separate publication~\cite{Meiser:2026}. 

\item The second solution is to compute the decay width in an operator basis where no such traces can emerge, as is the case for the Bern basis~\cite{Aebischer:2017gaw} as defined in the previous section. Due to the structure of the operators, built not to contain any $\gamma_5$ in the second current~\cite{Chetyrkin:1997gb}, the corresponding trace is free of $\gamma_5$. Hence, we can get at most one trace with $\gamma_5$, which can then be safely discarded.
However, this approach introduces operators containing long products of gamma matrices, which substantially increases the algebraic complexity of the resulting Dirac traces within the Feynman diagrams.
This is especially true for the insertions of evanescent operators, which we have to compute during the renormalization procedure as detailed in App.~\ref{app:renormalization}.
In our basis, they carry up to six gamma matrices per current making the traces cumbersome even for software packages like \texttt{FeynCalc}. 
We solved the issue by using \texttt{FormTracer}~\cite{Cyrol:2016zqb}, a \texttt{Mathematica} package, that lets one use a  \texttt{Mathematica} front end to pass the complicated traces to \texttt{FORM}, where the computation is done. 
Finally, the output is again given as a  \texttt{Mathematica} expression that can easily be interfaced with the rest of the computational pipeline. 
This is the solution that we will employ in this work. 

\end{enumerate}

Note that similar problems arise also for penguin insertions. However, in the Bern basis any penguin insertion is well-defined. In the interference of two open-penguin diagrams, there is only one Dirac trace, which raises no issues concerning $\gamma_5$. Exchanging one or two open-penguin diagrams for a closed-penguin one is also unproblematic in the Bern basis, since closed-penguin loops do not contain any $\gamma_5$. 
In more conventional bases, it is necessary to use Fierz transformations as well as flavor symmetries at NLO in order to compute some of the penguin-diagram contributions, especially when involving mixing with dipole operators. This will be further discussed in Ref.~\cite{Meiser:2026}.

%
\subsection{Loop Integration}
Once the Dirac traces have been computed, we have to perform one-loop integrations.
The usual UV divergences of the one-loop integrals are canceled by the appropriate counterterms. Some details on the renormalization procedure are provided in App.~\ref{app:renormalization}.
However, some of the contributions exhibit, apart from the UV divergences, additional IR divergences which complicate the phase-space integration. 
The $PP$, $DP$, and $DD$ topologies yield IR-finite contributions individually, and thus one can compute the phase-space integrals in four dimensions. Meanwhile, the three- and four-particle cuts in the NLO $CC$ insertions are individually IR-divergent. We regularize these IR divergences dimensionally, by computing the phase-space integrals in $d$ dimensions. 
However, since the IR divergences interact with the dimensionally-regulated loop integrals, it is important to keep them in their $d$-dimensional unexpanded form before computing the phase-space integrals. 
A full list of the unexpanded Passarino-Veltman loop integrals required for our calculation, which we encounter by using the standard reduction algorithm implemented in \texttt{FeynCalc}~\cite{Mertig:1990an,Shtabovenko:2016sxi,Shtabovenko:2020gxv,Shtabovenko:2023idz}, can be found in App.~\ref{app:Loops}.
Note that, as a cross-check, we computed the IR-finite contributions both in their completely unexpanded form using $d$-dimensional phase-space integrals, and by expanding the loop integrals in $\epsilon=2 - d/2$ and then integrating over the four-dimensional phase space, finding the same result in either case.

\subsection{Phase Space Integration}

After computing the IR-divergent integration kernels $\mathcal{K}^{(n)}_{ij}$, we need to perform the integration using the $d$-dimensional phase-space measure
\eqa{
    d\Pi_n &=& \prod_{i = 1}^n \frac{d^{d-1}p_i}{(2\pi)^{d-1} 2 E_i} \, (2\pi)^d \,  \delta^{(d)}\bigg(p_b- \sum_{i = 1}^n p_i\bigg) \ ,
}
with $(p_i,E_i)$ the momenta and energies of the final state particles. We will use the following notation,
\eqa{
I_{ij}^{(n)} \equiv \int d\Pi_n \, \mathcal{K}^{(n)}_{ij} \, .
}
%
\subsubsection{Three-particle Cuts}

For the three-particle cuts $b(p_b) \to q_1(p_1) \bar{q}_2(p_2) q_3(p_3)$, we find the following parametrization of the phase-space integral convenient, 
\eqa{
  I^{(3)}_{ij} &=& 2^{2-3 d} \left(\tilde{\mu}^2\right)^{4-d} \pi^{3-2 d} m_b^{2 d-6} \, \Omega_{d-2} \Omega_{d-1} \,  \int_0^1 dxdy \,  \frac{\left(x \bar{x} y^2
   \bar{y}\right)^{d/2}}{x^2 \bar{x}^2 y^3 \bar{y}^2} \,  \mathcal{K}^{(3)}_{ij} \, , 
}
with $\Omega_d = 2 \pi^{d/2}/\Gamma(d/2)$, $\tilde{\mu} = \mu \sqrt{e^{\gamma_E}/(4\pi)}$, $y = 2 E_3/m_b$ and $x = (E_2 + E_3 - m_b/2)/E_3$. 
After loop integration, the kernels $\mathcal{K}^{(3)}_{ij}$ are expressed in terms of $\Gamma$, $\beta$, and hypergeometric functions. 
Most of the phase-space integrals resulting from this parametrization can be evaluated analytically using the definitions of the $\beta$ and hypergeometric ${}_2F_1$ functions given in~Eq.~\ref{eq:BetaAndHyper}.
The remaining integrals have the form
\eqa{
\int_0^1 dx \, x^a \bar{x}^b \,  {}_2F_1(1,c+1,c+d+2;x) &=& \beta(1+a,1+b) \notag \\
& & \quad \times \,{}_3F_2(1,1+c,1+a,c+d+2,a+b+2,1) \, , \\
\int_0^1 dx dy \, \frac{x^a \bar{x}^b y^c \bar{y}^d}{1-xy} {}_2F_1(1,e,f,x y) \,  &=& I_4(a,b,c,d,e,f) \, .
}
The calculation of $I_4$ for the different parameters that appear in the calculation is quite involved and is delegated to App.~\ref{app:PS}.
With the integrals solved, the final result will be written in terms of $d$-dependent $\Gamma$ and $\beta$ functions as well as the hypergeometric functions ${}_2F_1$ and ${}_3F_2$. We expand them in $\epsilon = 2 - d/2$ using \texttt{HypExp} \cite{Huber:2007dx} and arrive at the final UV-finite but IR-divergent contribution from the three-particle cuts.

\subsubsection{Four-particle Cuts}
\label{sec:four-pt-cuts}

For the four-particle cuts $b(p_b) \to q_1(p_1) \bar{q}_2(p_2) q_3(p_3) g(p_g)$, we employ a similar approach as~Ref.~\cite{Huber:2014nna}, which is based on the form of the massless four-particle phase-space measure derived in~Ref.~\cite{Gehrmann-DeRidder:2003pne} in terms of kinematical invariants $\hat{s}_{ij} \equiv p_i \cdot p_j/m_b^2$ :
\begin{align}
    d\Pi_4 =& \tilde{\mu}^{12-3d} \, 2^{5-5d}\, \pi^{4-3d}\, m_b^{3d-8}\, (-\Delta_4)^{\frac{d-5}{2}} \, \delta(1 - \hat s_{12} - \hat s_{13} - \hat s_{1g} - \hat s_{23} - \hat s_{2g} - \hat s_{3g}) \notag \\
    & \times \Theta(-\Delta_4) \, d\Omega_{d-1} \, d\Omega_{d-2} \,d\Omega_{d-3} \, d\hat s_{12} \, d\hat s_{13} \, d\hat s_{1g} \, d\hat s_{23} \, d\hat s_{2g} \, d\hat s_{3g} \, ,
\end{align}
with the Gram determinant
\begin{align}
    \Delta_4 = \hat s_{12}^2 \hat s_{3g}^2 + \hat s_{13}^2 \hat s_{2g}^2 + \hat s_{1g}^2 \hat s_{23}^2 - 2 \hat s_{12} \hat s_{3g} \hat s_{13} \hat s_{2g} - 2  \hat s_{12} \hat s_{3g} \hat s_{1g} \hat s_{23} - 2\hat s_{13} \hat s_{2g} \hat s_{1g} \hat s_{23} \, . 
\end{align}
Performing the angular integration, inserting a fictitious integration variable $z$ along with a delta function, and performing a change of variables, we finally arrive at
\begin{align}
    I^{(4)}_{ij} 
    = &  N(d) \, 4^{d-4} \int_0^1 dz \, (z \bar{z})^{d-3}\, du \, dv \, dw \, dx \, \, (u \bar{u})^{\frac{d-5}{2}}  \, v^{d-3} (\bar{v} w \bar{w} x \bar{x})^{\frac{d-4}{2}} \, \mathcal{K}_{ij}^{(4)} \, ,
    \label{eq:4-pt-final}
\end{align}
where the two delta functions force
\begin{align}
\hat s_{13}\to z - \hat s_{12} - \hat s_{23} \, , \quad \hat s_{2g} \to \bar{z} - \hat s_{1g} - \hat s_{3g} \, .
\end{align}
The remaining kinematic invariants are transformed as follows
\begin{align}
    \hat s_{12} = z v w \, , \quad \hat s_{1g} = \bar{z} v x \, , \quad \hat s_{3g} = \bar{z} \bar{v} \, , \quad \hat s_{23}  = 4zu\sqrt{\bar{v}w\bar{w}x\bar{x}} + a^- \, .
\end{align}
The transformation of $\hat{s}_{23}$ involves the quantity $a^-$ given by
\begin{align}
   a^- = z(\bar{v}wx+\bar{x}\bar{w}-2\sqrt{\bar{v}w\bar{w}x\bar{x}}) \, .
\end{align}
Finally, the prefactor is defined as
\begin{align}
    N(d) = \tilde{\mu}^{12-3d} \, 2^{5-5d}\, \pi^{4-3d}\, m_b^{3d-8}\,\Omega_{d-1} \, \Omega_{d-2} \, \Omega_{d-3} \, .
\end{align}
This form of the four-particle phase-space is derived in more detail in Ref.~\cite{Huber:2014nna}.
The upshot is that, with the variables of~Eq.~(\ref{eq:4-pt-final}), all the terms of the integration kernel are of the form
\eqa{
\mathcal{K}_{ij}^{(4)} \supset x^a \bar{x}^b w^c \bar{w}^d v^e \bar{v}^f u^g \bar{u}^h z^i \bar{z}^j \, , 
}
with some parameters $a, \, \dots \, ,j$. Integrating this over the unit cube, yields $\beta$ functions that can easily be expanded in $\epsilon$. The resulting IR-divergent results will compensate for the remaining IR divergences of the three-particle cuts.

\section{Results}
\label{sec:results}

We present here the main results of our calculation, which are the matrices describing the partonic decay rate in terms of the different Wilson coefficients, separated in the four different cases of interest, and further divided in terms of the diagram topologies involved in their calculation, namely: current-current ($CC$) insertions on the one hand; and a combination of penguin-penguin ($PP$) insertions, penguin-dipole ($PD$) and dipole-dipole ($DD$) insertions on the other hand, overall labeled as $P$. The resulting picture looks like:
\begin{eqnarray}
    \text{Case 1:} \qquad & \Gamma(b \to d \bar u u) =& \Gamma_0\sum_{i,j} \Big( \mathcal{C}^{dbuu\dagger}_i \mathcal{C}^{dbuu}_j + \mathcal{C}^{dbuu\dagger}_{i'} \mathcal{C}^{dbuu}_{j'} \Big) \, G^{ij}_{CC}[d \bar u u] \notag \\
    & & +  \Gamma_0 \sum_{i,j} \Big( \mathcal{C}^{dbX\dagger}_i \mathcal{C}^{dbX}_j + \mathcal{C}^{dbX\dagger}_{i'}\mathcal{C}^{dbX}_{j'} \Big) \, G^{ij}_{P}[d \bar u u] \, , \label{eq:GMat_1} \\
    \text{Case 2:}  \qquad & \Gamma(b \to d \bar s d) =& \Gamma_0\sum_{i,j} \Big( \mathcal{C}^{dbds\dagger}_{i} \mathcal{C}^{dbds}_{j} + \mathcal{C}^{dbds\dagger}_{i'} \mathcal{C}^{dbds}_{j'} \Big) \, G^{ij}_{CC}[d \bar s d] \, , \label{eq:GMat_2} \\
    \text{Case 3:}  \qquad & \Gamma(b \to d \bar d d) =& \Gamma_0\sum_{i,j} \Big( \mathcal{C}^{dbdd\dagger}_i \mathcal{C}^{dbdd}_j + \mathcal{C}^{dbdd\dagger}_{i'} \mathcal{C}^{dbdd}_{j'} \Big) \, G^{ij}_{CC}[d \bar d d] \notag \\*
    & & +  \Gamma_0 \sum_{i,j} \Big( \mathcal{C}^{dbX\dagger}_i \mathcal{C}^{dbX}_j + \mathcal{C}^{dbX\dagger}_{i'}\mathcal{C}^{dbX}_{j'} \Big) \, G^{ij}_{P}[d \bar d d] \, , \label{eq:GMat_3} \\
    \text{Case 4:}  \qquad & \Gamma(b \to d g) =& \Gamma_0 \sum_{i,j} \Big( \mathcal{C}^{dbX\dagger}_i \mathcal{C}^{dbX}_j + \mathcal{C}^{dbX\dagger}_{i'}\mathcal{C}^{dbX}_{j'} \Big) \, G^{ij}_{P}[d g] \, , \label{eq:GMat_4}
\end{eqnarray}
with $\Gamma_0 = G_F^2 m_b^5/(192 \pi^3)$. Furthermore, the current-current matrices $\hat{G}_{CC}[d\bar{q} q']$ allow for a perturbative expansion in $\tilde{\alpha}_s \equiv g_s^2/(4 \pi)^2$,
\eqa{
\label{eq:Gexpansion}
\hat{G}_{CC}[d\bar{q} q'] = \hat{G}_{CC}^{(0)}[d\bar{q} q'] + \tilde{\alpha}_s \, \hat{G}_{CC}^{(1)}[d\bar{q} q'] + \mathcal{O}(\tilde{\alpha}_s^2) \, , 
}
while the $\hat{G}_{P}[\dots]$ start already at $\mathcal{O}(\tilde{\alpha}_s)$, in accordance with the power-counting assumption explained below~Eq.~(\ref{eq:OP_dipoles}). Note that the NLO matrices will also be functions of $L_{b(c)} \equiv \log\mu/m_{b(c)}$.

Our results include all previously introduced matrices, computed to NLO in the expansion in~$\tilde{\alpha}_s$. We have also cross-checked all results with an independent calculation based on a more standard operator basis, which uses the framework of changes of basis at NLO both to avoid the problematic traces of $\gamma_5$ and to relate the results to the Bern basis, to be published elsewhere~\cite{Meiser:2026}.
In addition to the results presented in the text below, we provide the  LO and NLO results in digital format as an ancillary \texttt{Mathematica} file ``\texttt{BlifetimeNLO.m}''. 

\subsection{Case 1 ($b \to d \bar{u} u$)}

Here we present the full set of LO and NLO contributions to the decay rate of the parton-level processes enclosed as Case 1, separated in current-current and penguin-like ($PP$ and $PD$) insertions. Note that while~Eq.~(\ref{eq:GMat_1}) introduces $\hat{G}_{CC}[d \bar u u]$ and $\hat{G}_{P}[d \bar u u]$ as concerning different sets of Wilson coefficients, $\big\{\mathcal{C}^{dbuu}_{i^{(\prime)}}\big\}$ is a subset of $\big\{\mathcal{C}^{dbX}_{i^{(\prime)}}\big\}$, and thus both contributions can eventually be combined into a single $41\times41$ matrix. 

We also remark that, while we focus on the specific case of $b \to d \bar{u} u$ for definiteness, the other possible flavor combinations can be readily derived from these results.

\subsubsection{Current-Current Insertions}
\label{sec:CCInsertions}

The corresponding $\hat G$ matrices read, at leading order,
\eqa{
\hat{G}_{CC}^{(0)}[d \bar{u} u] = \left(
\begin{array}{cc}
 \hat{G}_{1-4}^{(0)}[d \bar{u} u] & 0_{4 \times 6} \\
 0_{6 \times 4} & \hat{G}_{5-10}^{(0)}[d \bar{u} u]
\end{array}
\right) \, ,
}
with the blocks being
\eqa{
\hat{G}_{1-4}^{(0)}[d \bar{u} u] =  \left(
\begin{array}{cccc}
 6 & 0 & 60 & 0  \\
 0 & \frac{4}{3} & 0 & \frac{40}{3} \\
 60 & 0 & 816 & 0 \\
 0 & \frac{40}{3} & 0 & \frac{544}{3} \\
\end{array}
\right) \, ,\quad
\hat{G}_{5-10}^{(0)}[d \bar{u} u] = \left(
\begin{array}{cccccc}
 \frac{3}{2} & 0 & 0 & 0 & 60 & 0 \\
 0 & \frac{1}{3} & 0 & 0 & 0 & \frac{40}{3} \\
 0 & 0 & 36 & 0 & -576 & 0 \\
 0 & 0 & 0 & 8 & 0 & -128 \\
 60 & 0 & -576 & 0 & 12480 & 0 \\
 0 & \frac{40}{3} & 0 & -128 & 0 & \frac{8320}{3}
\end{array}
\right) \, .
}
The LO result agrees with the $m_c \to 0$ limit of the result for $b\to d(s) \bar c u$ in Refs.~\cite{Meiser:2024zea,Lang:2025ios}.
The NLO result has the same block structure as the LO result:
\eqa{
\hat{G}_{CC}^{(1)}[d \bar{u} u] = \left(
\begin{array}{cc}
 \hat{G}_{1-4}^{(1)}[d \bar{u}u] & 0_{4 \times 6} \\
 0_{6 \times 4} & \hat{G}_{5-10}^{(1)}[d \bar{u}u]
\end{array}
\right) \, ,
}
with the blocks being
\eqa{
& & \hat{G}_{1-4}^{(1)}[d \bar{u} u] = \left(
\begin{array}{cccc}
 124-16 \pi ^2 & -16 & 1240-160 \pi ^2 & -392 \\
 -16 & \frac{557}{9}-\frac{32 \pi ^2}{9} & -392 & \frac{4700}{9}-\frac{320 \pi ^2}{9} \\
 1240-160 \pi ^2 & -392 & 21472-2176 \pi ^2 & -6048 \\
 -392 & \frac{4700}{9}-\frac{320 \pi ^2}{9} & -6048 & \frac{54896}{9}-\frac{4352 \pi ^2}{9} \\
\end{array}
\right) \notag \\[6pt]
& & \qquad \qquad \qquad \qquad + L_b \left(
\begin{array}{cccc}
 0 & 0 & 0 & -192 \\
 0 & 24 & -192 & 160 \\
 0 & -192 & 0 & -3840 \\
 -192 & 160 & -3840 & 1664 \\
\end{array}
\right) \, , \\[8pt]
& & \hat{G}_{5-10}^{(1)}[d \bar{u} u] = \notag \\*
& & \quad \left(
\begin{array}{cccccc}
 89-4 \pi ^2 & 0 & 0 & \frac{164}{3} & 3560-160 \pi ^2 & -\frac{2288}{3} \\
 0 & \frac{149}{18}-\frac{8 \pi ^2}{9} & \frac{164}{3} & \frac{205}{9} & -\frac{2288}{3} & \frac{40}{3}-\frac{320 \pi ^2}{9} \\
 0 & \frac{164}{3} & 536-96 \pi ^2 & -192 & 1536 \pi ^2-13184 & \frac{16256}{3} \\
 \frac{164}{3} & \frac{205}{9} & -192 & \frac{3268}{9}-\frac{64 \pi ^2}{3} & \frac{16256}{3} & \frac{1024 \pi ^2}{3}-4128 \\
 3560-160 \pi ^2 & -\frac{2288}{3} & 1536 \pi ^2-13184 & \frac{16256}{3} & 515200-33280 \pi ^2 & -\frac{325120}{3} \\
 -\frac{2288}{3} & \frac{40}{3}-\frac{320 \pi ^2}{9} & \frac{16256}{3} & \frac{1024 \pi ^2}{3}-4128 & -\frac{325120}{3} & \frac{174400}{3}-\frac{66560 \pi ^2}{9}
\end{array}
\right) \notag \\[6pt]
& & \qquad + L_b 
\left(
\begin{array}{cccccc}
 48 & 0 & 0 & 32 & 1920 & -512 \\
 0 & -\frac{4}{3} & 32 & \frac{40}{3} & -512 & -\frac{800}{3} \\
 0 & 32 & -384 & 0 & 6144 & 2048 \\
 32 & \frac{40}{3} & 0 & \frac{608}{3} & 2048 & -\frac{7168}{3} \\
 1920 & -512 & 6144 & 2048 & 6144 & -65536 \\
 -512 & -\frac{800}{3} & 2048 & -\frac{7168}{3} & -65536 & \frac{65024}{3} \\
\end{array}
\right) \, .
}

\bigskip

\subsubsection{Penguin-Penguin and  Penguin-Dipole Insertions}

$PP$ and $DP$ insertions do not contribute at LO and so there is no such term in the expansion of $\hat{G}_{P}$. The block structure at NLO is
\eqa{
\hat{G}_{P}^{(1)}[d\bar{u}u] =  \left(\begin{array}{cccccc}
    \hat{G}_{PP}^{uu}[d\bar{u}u] & \hat{G}_{PP}^{us}[d\bar{u}u] & \hat{G}_{PP}^{uc}[d\bar{u}u] & \hat{G}_{PP}^{ud}[d\bar{u}u] & \hat{G}_{PP}^{ub}[d\bar{u}u] & \hat{G}_{PD}^{ug}[d\bar{u}u] \\
    \hat{G}_{PP}^{su}[d\bar{u}u] & 0 & 0 & 0 & 0 & 0 \\
    \hat{G}_{PP}^{cu}[d\bar{u}u] & 0 & 0 & 0 & 0 & 0 \\
    \hat{G}_{PP}^{du}[d\bar{u}u] & 0 & 0 & 0 & 0 & 0 \\
    \hat{G}_{PP}^{bu}[d\bar{u}u] & 0 & 0 & 0 & 0 & 0 \\
    \hat{G}_{DP}^{gu}[d\bar{u}u] & 0 & 0 & 0 & 0 & 0
    \label{eq:lifetimePPInsertions}
\end{array}
\right) \, .
}
The different sub-blocks read: 
\eqa{
& & \hat{G}_{PP}^{us}[d\bar{u}u] = \hat{G}_{PP}^{su \, \dagger}[d\bar{u}u] \notag \\*[8pt]
& & = -\frac{1}{9} \left(
\begin{array}{cccccccccc}
 0 & 0 & 0 & 0 & 0_{1 \times 6} \\
 0 & 26 + 16 L_b + 8 i \pi & 0 & 212 + 160 L_b + 80 i \pi & 0_{1 \times 6}\\
 0 & 0 & 0 & 0  & 0_{1 \times 6}\\
 0 & 260 + 160 L_b + 80 i \pi & 0 & 2120 + 1600 L_b + 800 i \pi & 0_{1 \times 6} \\
 0_{6 \times 1} & 0_{6 \times 1} & 0_{6 \times 1} & 0_{6 \times 1} & 0_{6 \times 6} \\
\end{array}
\right) \, , \\[8pt]
& & \hat{G}_{PP}^{uc \, \dagger}[d\bar{u}u] = \hat{G}_{PP}^{cu}[d\bar{u}u] \notag \\*[8pt]
& & =  \left(
\begin{array}{ccccc}
 0 & 0 & 0 & 0 & 0_{1 \times 6} \\
 0 & f(z_c) - \frac{16}{9}L_b & 0 & 10 f(z_c) - \frac{160}{9}L_b & 0_{1 \times 6} \\
 0 & 0 & 0 & 0 & 0_{1 \times 6} \\
 0 & 10 f(z_c) + \frac{16}{3} - \frac{160}{9}L_b & 0 & 100 f(z_c) + \frac{160}{3} - \frac{1600}{9}L_b & 0_{1 \times 6} \\
 0_{3 \times 1} & 0_{3 \times 1} & 0_{3 \times 1} & 0_{3 \times 1} & 0_{3 \times 6} \\
 0 & z_c \left( g(z_c) - \frac{64}{3} L_b \right) & 0 & 10 z_c \left( g(z_c) - \frac{64}{3} L_b \right) & 0_{1 \times 6} \\
 0 & 0 & 0 & 0 & 0_{1 \times 6} \\
 0 & z_c \left( -16 g(z_c) - 128 + \frac{1024}{3} L_b \right) & 0 & z_c \left( -160 g(z_c) - 1280 + \frac{10240}{3} L_b \right) & 0_{1 \times 6}
\end{array} \right) \, , \\[8pt]
& & \hat{G}_{PP}^{ud \, \dagger}[d\bar{u}u] = \hat{G}_{PP}^{du}[d\bar{u}u] \notag \\*[8pt] 
&& = -\frac{1}{9} \left(
\begin{array}{ccccc}
 0 & 18 + 16 L_b - 8 i \pi & 0 & 180 + 160 L_b - 80 i \pi & 0_{1 \times 6} \\
 0 & 256(1 + L_b) - 128 i \pi & 0 & 2560(1 + L_b) - 1280 i \pi & 0_{1 \times 6} \\
 0 & -(13 + 8 L_b - 4 i \pi) & 0 & -(130 + 80 L_b - 40 i \pi) & 0_{1 \times 6} \\
 0 & 24 & 0 & 240 & 0_{1 \times 6} \\
 0 & -(720 + 128 L_b - 64 i \pi) & 0 & -(7200 + 1280 L_b - 640 i \pi) & 0_{1 \times 6}
\end{array} \right) \, ,
}
\eqa{
& & \hat{G}_{PP}^{ub \, \dagger}[d\bar{u}u] = \hat{G}_{PP}^{bu}[d\bar{u}u] \notag \\*[8pt]
& & = \left(
\begin{array}{ccccc}
 0 & \frac{82}{3} - 8 \sqrt{3} \pi + \frac{32 \pi^2}{27} - \frac{16}{9} L_b & 0 & \frac{820}{3} - 80 \sqrt{3} \pi + \frac{320 \pi^2}{27} - \frac{160}{9} L_b & 0_{1 \times 6} \\
 0 & \frac{3776}{9} - 128 \sqrt{3} \pi + \frac{512 \pi^2}{27} - \frac{256}{9} L_b & 0 & \frac{37760}{9} - 1280 \sqrt{3} \pi + \frac{5120 \pi^2}{27} - \frac{2560}{9} L_b & 0_{1 \times 6} \\
 0 & -7 + 4 \sqrt{3} \pi - \frac{40 \pi^2}{27} + \frac{56}{9} L_b & 0 & -70 + 40 \sqrt{3} \pi - \frac{400 \pi^2}{27} + \frac{560}{9} L_b & 0_{1 \times 6} \\
 0 & \frac{32 \pi^2}{9} - \frac{296}{9} - \frac{64}{3} L_b & 0 & \frac{320 \pi^2}{9} - \frac{2960}{9} - \frac{640}{3} L_b & 0_{1 \times 6} \\
 0 & \frac{4624}{9} + 64 \sqrt{3} \pi - \frac{3328 \pi^2}{27} + \frac{6272}{9} L_b & 0 & \frac{46240}{9} + 640 \sqrt{3} \pi - \frac{33280 \pi^2}{27} + \frac{62720}{9} L_b & 0_{1 \times 6}
\end{array} \right) \, , \\[8pt]
& & \hat{G}_{DP}^{gu}[d\bar{u}u] = \hat{G}_{PD}^{ug \, \dagger}[d\bar{u}u] = \frac{1}{\tilde{\alpha}_s} \left (
\begin{array}{ccccc}
    0 & -\frac{16}{3} & 0 & -\frac{160}{3} & 0_{1 \times 6} 
\end{array} \right ) \, ,
}
with
\eqa{
f(z_c) &=& \frac{2}{9} \left(12 z_c^2 \left(12 \pi ^2 z_c^6+\left(12-8 \pi ^2\right) z_c^4-7 z_c^2+8\right) + 8 \log z_c -13\right) \notag \\*
& & +\frac{2}{9} \left (4 i \pi \Delta_c \left(1 - 36  z_c^6 + 18 z_c^4 - 10  z_c^2 \right ) \right ) \notag \\*
& & +\frac{8}{9} 
   \left(\Delta_c  \left(-36 z_c^6+18 z_c^4-10 z_c^2+1\right)-24 i \pi  \left(3 z_c^2-2\right) z_c^6\right)\Lambda_c \notag \\*
& & + \left(\frac{64 z_c^6}{3}-32 z_c^8\right) \Lambda_c^2 \, ,  \\[8pt]
g(z_c) &=& \frac{16}{9} \left(36 \pi ^2z_c^6-36 \left(\pi ^2-1\right)z_c^4+3 \left(6 \pi ^2-11\right)z_c^2+12 \log (z_c)-23\right) \notag \\*
& & -\frac{32}{3} \left ( i \pi \Delta_c \left(-1  +6  z_c^4 -5 z_c^2 \right) \right ) \notag \\*
& &+\frac{32}{3}  \left(\Delta_c  \left(-6 z_c^4+5
   z_c^2+1\right) - 6 i \pi  \left(2 z_c^6-2 z_c^4+z_c^2\right)\right)\Lambda_c \notag \\*
& & -32 \left(2 z_c^6-2 z_c^4+z_c^2\right) \Lambda_c^2 \, ,
}
where we used the shorthand
\eqa{
z_c \equiv \frac{m_c}{m_b} \, , \qquad \Delta_c \equiv \sqrt{1-4 z_c^2} \, , \qquad \Lambda_c \equiv \log \frac{1+ \Delta_c}{1-\Delta_c} \, .
}
Finally, we have
\eqa{
\hat{G}_{PP}^{uu}[d\bar{u}u] = \hat{G}_{PP}^{us}[d\bar{u}u]+ \hat{G}_{PP}^{su}[d\bar{u}u] \, .
}
Let us note here that the dipole contributions carry a factor of $1/\tilde{\alpha}_s$, which might (mis)lead to the conclusion that they are enhanced. However, just as argued below~Eq.~(\ref{eq:OP_dipoles}), the dipole Wilson coefficients in the LEFT/WET receive tipically a suppression of at least $1/16\pi^2$ in their matching onto the SMEFT (and in the latter's matching onto further UV completions), which will cancel the apparent enhancement, thus rendering the dipole contributions a true NLO effect\footnote{Even assuming the minimum suppression of $1/16\pi^2$, the cancellation with the apparent $1/\tilde{\alpha}_s$ enhancement yields a factor of $1/g_s^2<1$.}.

\subsection{Case 2 ($b \to d \bar{s} d$)}
For $b \to d \bar s d$, there are only $CC$ topologies. The tree-level contribution reads
\eqa{
\hat{G}^{(0)}_{CC}[d\bar{s}d] =\left(
\begin{array}{ccccc}
 7 & 76 & -\frac{1}{2} & 0 & -8 \\
 76 & 1072 & -2 & 0 & -32 \\
 -\frac{1}{2} & -2 & \frac{11}{8} & -\frac{3}{2} & 76 \\
 0 & 0 & -\frac{3}{2} & 42 & -768 \\
 -8 & -32 & 76 & -768 & 16576 \\
\end{array}
\right) \, .
}
At NLO, we find
\eqa{
\hat{G}^{(1)}_{CC}[d\bar{s}d] &=& \left(
\begin{array}{ccccc}
 \frac{280}{3}-\frac{56 \pi ^2}{3} & \frac{2536}{3}-\frac{608 \pi ^2}{3} & \frac{4 \pi ^2}{3}-\frac{205}{12} & -\frac{269}{3} & \frac{6136}{3}+\frac{64 \pi ^2}{3} \\
 \frac{2536}{3}-\frac{608 \pi ^2}{3} & \frac{50944}{3}-\frac{8576 \pi ^2}{3} & \frac{16 \pi ^2}{3}-\frac{73}{3} & -\frac{1076}{3} & \frac{26656}{3}+\frac{256 \pi ^2}{3}
   \\
 \frac{4 \pi ^2}{3}-\frac{205}{12} & \frac{16 \pi ^2}{3}-\frac{73}{3} & 64-\frac{11 \pi ^2}{3} & 4 \pi ^2-\frac{227}{6} & \frac{8194}{3}-\frac{608 \pi ^2}{3} \\
 -\frac{269}{3} & -\frac{1076}{3} & 4 \pi ^2-\frac{227}{6} & \frac{2068}{3}-112 \pi ^2 & 2048 \pi ^2-\frac{65960}{3} \\
 \frac{6136}{3}+\frac{64 \pi ^2}{3} & \frac{26656}{3}+\frac{256 \pi ^2}{3} & \frac{8194}{3}-\frac{608 \pi ^2}{3} & 2048 \pi ^2-\frac{65960}{3} & 676480-\frac{132608 \pi
   ^2}{3} \\
\end{array}
\right)  \notag \\[8pt]
& & + L_b
\left(
\begin{array}{ccccc}
 -32 & -512 & -16 & 0 & -256 \\
 -512 & -8192 & -64 & 0 & -1024 \\
 -16 & -64 & 36 & -16 & 1408 \\
 0 & 0 & -16 & -832 & 12288 \\
 -256 & -1024 & 1408 & 12288 & -124928 \\
\end{array}
\right) \, .
}
\subsection{Case 3 ($b \to d\bar{d}d$)}
For $b \to d \bar{d}d$, the $CC$ insertions yield exactly the same result as for the Case 2 channels:
\eqa{
\hat{G}_{CC}^{(0)}[d\bar d d] = \hat{G}_{CC}^{(0)}[d\bar s d] \, , \quad \hat{G}_{CC}^{(1)}[d\bar d d] = \hat{G}_{CC}^{(1)}[d\bar s d] \, . 
}
The $PP$ and $DP$ insertions, starting at NLO, have the following block structure:
\eqa{
\hat{G}_{P}^{(1)}[d\bar{d}d] =  \left(\begin{array}{cccccc}
    0 & 0 & 0 & \hat{G}_{PP}^{ud}[d\bar{d}d] & 0 & 0\\
    0 & 0 & 0 & \hat{G}_{PP}^{sd}[d\bar{d}d] & 0 & 0\\
    0 & 0 & 0 & \hat{G}_{PP}^{cd}[d\bar{d}d] & 0 & 0\\
    \hat{G}_{PP}^{du}[d\bar{d}d] & \hat{G}_{PP}^{ds}[d\bar{d}d] & \hat{G}_{PP}^{dc}[d\bar{d}d] & \hat{G}_{PP}^{dd}[d\bar{d}d] & \hat{G}_{PP}^{db}[d\bar{d}d] & \hat{G}_{PD}^{dg}[d\bar{d}d]  \\
    0 & 0 & 0 & \hat{G}_{PP}^{bd}[d\bar{d}d] & 0 & 0 \\
    0 & 0 & 0 & \hat{G}_{DP}^{gd}[d\bar{d}d] & 0 & 0
    \label{eq:b->ddd:PP}
\end{array}
\right) \, ,
}
with blocks reading
\eqa{
\hat{G}_{PP}^{ud}[d\bar{d}d] &=& \hat{G}_{PP}^{du \, \dagger}[d\bar{d}d] = \hat{G}_{PP}^{sd}[d\bar{d}d] = \hat{G}_{PP}^{ds \, \dagger}[d\bar{d}d] \notag  \\*[8pt]
&=& \left(
\begin{array}{ccccc}
 0 & 0 & 0 & 0 & 0 \\
 -\frac{29}{9}+\frac{8 i \pi }{9} & -\frac{476}{9}+\frac{128 i \pi }{9} & \frac{13}{18}-\frac{4 i \pi }{9} & \frac{20}{3} & -\frac{1288}{9}-\frac{64 i \pi }{9} \\
 0 & 0 & 0 & 0 & 0 \\
 -\frac{242}{9}+\frac{80 i \pi }{9} & -\frac{3992}{9}+\frac{1280 i \pi }{9} & \frac{41}{9}-\frac{40 i \pi }{9} & \frac{200}{3} & -\frac{13264}{9}-\frac{640 i \pi }{9} \\
 0_{6 \times 1} & 0_{6 \times 1} & 0_{6 \times 1} & 0_{6 \times 1} & 0_{6 \times 1} 
\end{array}
\right) \notag \\*[8pt]
& & + L_b
\left(
\begin{array}{ccccc}
 0 & 0 & 0 & 0 & 0 \\
 -\frac{16}{9} & -\frac{256}{9} & \frac{8}{9} & 0 & \frac{128}{9} \\
 0 & 0 & 0 & 0 & 0 \\
 -\frac{160}{9} & -\frac{2560}{9} & \frac{80}{9} & 0 & \frac{1280}{9} \\
 0_{6 \times 1} & 0_{6 \times 1} & 0_{6 \times 1} & 0_{6 \times 1} & 0_{6 \times 1}  
\end{array}
\right) \, , \\[8pt]
\hat{G}_{PP}^{cd}[d\bar{d}d] &=& \hat{G}_{PP}^{dc \, \dagger}[d\bar{d}d] \notag \\*[8pt]
&=& \left(
\begin{array}{ccccc}
 0 & 0 & 0 & 0 & 0 \\
 f(z_c)-\frac{1}{3} & 16 f(z_c)-\frac{20}{3} & \frac{1}{18} \left(-9 f(z_c)-13\right) & \frac{20}{3} & -\frac{8}{9} \left(9
   f(z_c)+187\right) \\
 0 & 0 & 0 & 0 & 0 \\
 10 f(z_c)+2 & \frac{8}{3} \left(60 f(z_c)+7\right) & -5 f(z_c)-\frac{89}{9} & \frac{200}{3} & -\frac{16}{9} \left(45 f(z_c)+959\right) \\
 0_{3\times1} &  0_{3\times1} &  0_{3\times1} &  0_{3\times1} &  0_{3\times1} \\
 z_c g(z_c) & 16 z_c g(z_c) & -\frac{1}{2} z_c g(z_c) & 0 & -8 z_c g(z_c) \\
 0 & 0 & 0 & 0 & 0 \\
 -16 z_c \left(g(z_c)+8\right) & -256 z_c \left(g(z_c)+8\right) & 8 z_c \left(g(z_c)+8\right) & 0 & 128 z_c
   \left(g(z_c)+8\right) \\
\end{array}
\right) \notag \\[8pt]
& & + L_b
\left(
\begin{array}{ccccc}
 0 & 0 & 0 & 0 & 0 \\
 -\frac{16}{9} & -\frac{256}{9} & \frac{8}{9} & 0 & \frac{128}{9} \\
 0 & 0 & 0 & 0 & 0 \\
 -\frac{160}{9} & -\frac{2560}{9} & \frac{80}{9} & 0 & \frac{1280}{9} \\
 0_{3\times1} &  0_{3\times1} &  0_{3\times1} &  0_{3\times1} &  0_{3\times1} \\
 -\frac{64 z_c}{3} & -\frac{1024 z_c}{3} & \frac{32 z_c}{3} & 0 & \frac{512 z_c}{3} \\
 0 & 0 & 0 & 0 & 0 \\
 \frac{1024 z_c}{3} & \frac{16384 z_c}{3} & -\frac{512 z_c}{3} & 0 & -\frac{8192 z_c}{3} \\
\end{array}
\right)  \, , \\[8pt]
\hat{G}_{PP}^{dd}[d\bar{d}d] &=& 
\left(
\begin{array}{ccccc}
 -\frac{14}{3} & -\frac{652}{9} & \frac{17}{9} & 4 & -\frac{608}{9} \\
 -\frac{652}{9} & -\frac{10112}{9} & \frac{262}{9} & 64 & -\frac{3296}{3} \\
 \frac{17}{9} & \frac{262}{9} & -\frac{13}{18} & -2 & \frac{112}{3} \\
 4 & 64 & -2 & 0 & -32 \\
 -\frac{608}{9} & -\frac{3296}{3} & \frac{112}{3} & -32 & \frac{12416}{9}
\end{array}
\right)
+ L_b
\left(
\begin{array}{ccccc}
 -\frac{32}{9} & -\frac{512}{9} & \frac{16}{9} & 0 & \frac{256}{9} \\
 -\frac{512}{9} & -\frac{8192}{9} & \frac{256}{9} & 0 & \frac{4096}{9} \\
 \frac{16}{9} & \frac{256}{9} & -\frac{8}{9} & 0 & -\frac{128}{9} \\
 0 & 0 & 0 & 0 & 0 \\
 \frac{256}{9} & \frac{4096}{9} & -\frac{128}{9} & 0 & -\frac{2048}{9} \\
\end{array}
\right) \, , \\[8pt]
\hat{G}_{PP}^{bd}[d\bar{d}d] &=& \hat{G}_{PP}^{db \, \dagger}[d\bar{d}d] \notag \\[8pt]
& = &
\left(
\begin{array}{ccc}
 27-8 \sqrt{3} \pi +\frac{32 \pi ^2}{27} & \frac{1292}{3}-128 \sqrt{3} \pi +\frac{512 \pi ^2}{27} & -\frac{259}{18}+4 \sqrt{3} \pi -\frac{16 \pi ^2}{27} \\
 \frac{3728}{9}-128 \sqrt{3} \pi +\frac{512 \pi ^2}{27} & \frac{59456}{9}-2048 \sqrt{3} \pi +\frac{8192 \pi ^2}{27} & -\frac{664}{3}+64 \sqrt{3} \pi -\frac{256 \pi ^2}{27}
   \\
 -\frac{41}{6}+4 \sqrt{3} \pi -\frac{40 \pi ^2}{27} & -\frac{326}{3}+64 \sqrt{3} \pi -\frac{640 \pi ^2}{27} & \frac{139}{36}-2 \sqrt{3} \pi +\frac{20 \pi ^2}{27} \\
 \frac{32 \pi ^2}{9}-\frac{296}{9} & \frac{512 \pi ^2}{9}-\frac{4736}{9} & \frac{148}{9}-\frac{16 \pi ^2}{9} \\
 \frac{4648}{9}+64 \sqrt{3} \pi -\frac{3328 \pi ^2}{27} & \frac{74464}{9}+1024 \sqrt{3} \pi -\frac{53248 \pi ^2}{27} & -\frac{2260}{9}-32 \sqrt{3} \pi +\frac{1664 \pi
   ^2}{27} \\
\end{array}
\right . \notag \\[8pt]
& & \quad  \left .
\begin{array}{cc}
 \frac{20}{3} & -\frac{3464}{9}+64 \sqrt{3} \pi -\frac{256 \pi ^2}{27} \\
 \frac{320}{3} & -6016+1024 \sqrt{3} \pi -\frac{4096 \pi ^2}{27} \\
 -\frac{10}{3} & \frac{1252}{9}-32 \sqrt{3} \pi +\frac{320 \pi ^2}{27} \\
 0 & \frac{2368}{9}-\frac{256 \pi ^2}{9} \\
 -\frac{160}{3} & -\frac{25024}{9}-512 \sqrt{3} \pi +\frac{26624 \pi ^2}{27} \\
\end{array}
\right) + L_b
\left(
\begin{array}{ccccc}
 -\frac{16}{9} & -\frac{256}{9} & \frac{8}{9} & 0 & \frac{128}{9} \\
 -\frac{256}{9} & -\frac{4096}{9} & \frac{128}{9} & 0 & \frac{2048}{9} \\
 \frac{56}{9} & \frac{896}{9} & -\frac{28}{9} & 0 & -\frac{448}{9} \\
 -\frac{64}{3} & -\frac{1024}{3} & \frac{32}{3} & 0 & \frac{512}{3} \\
 \frac{6272}{9} & \frac{100352}{9} & -\frac{3136}{9} & 0 & -\frac{50176}{9} \\
\end{array}
\right) \, ,
\\[8pt]
\hat{G}_{DP}^{gd}[d\bar{d}d] &=& \hat{G}_{PD}^{dg \, \dagger}[d\bar{d}d] = \frac{1}{\tilde{\alpha}_s}\left(
\begin{array}{ccccc}
    -\frac{16}{3} &-\frac{256}{3} & \frac{8}{3} & 0 & \frac{128}{3}
\end{array} 
\right) \, .
}
Just as in the $b\to d \bar u u$ case, while $\hat{G}_{CC}[d \bar d d]$ is a smaller matrix concerning only the 5 Wilson coefficients $\big\{\mathcal{C}^{dbdd}_{i^{(\prime)}}\big\}$, the latter is a subset of $\big\{\mathcal{C}^{dbX}_{i^{(\prime)}} \big\}$, and therefore both contributions can eventually be recombined into a single $41\times41$ matrix.

\subsection{Case 4 ($b \to dg$)}
Here we present the results for the contribution to the $b$-quark decay rate from two-particle cuts, $b \to d g$. In our limit $m_u = m_d = m_s = 0$, the only non-zero contributions are those coming from charm- and bottom-loop $PP$ insertions, $DD$ insertions, and the interference $PD$. Hence, the block structure of the matrix is
\eqa{
\hat{G}_P^{(1)}[dg] =  \left(\begin{array}{cccccc}
   0_{20 \times 20} & 0_{20 \times 10} & 0_{20 \times 5} & 0_{20 \times 5} & 0_{20 \times 1} \\
    0_{10 \times 20} & \hat{G}_{dg}^{cc} & 0_{10 \times 5} & \hat{G}_{dg}^{cb} & \hat{G}_{dg}^{cG} \\
   0_{5 \times 20} & 0_{5 \times 10} & 0_{5 \times 5} & 0_{5 \times 5} & 0_{5 \times 1} \\
   0_{5 \times 20} & \hat{G}_{dg}^{bc} &  0_{5 \times 5} & \hat{G}_{dg}^{bb} & \hat{G}_{dg}^{bG} \\
   0_{1 \times 20} & \hat{G}_{dg}^{Gc} & 0_{1 \times 5} & \hat{G}_{dg}^{Gb} & \hat{G}_{dg}^{GG}
\end{array}
\right) \, .
}
The sub-blocks of this matrix read
\eqa{
\hat{G}_{dg}^{cc} &=& \left(
\begin{array}{cccc}
 0_{7 \times 7} & 0_{7 \times 1} & 0_{7 \times 1} & 0_{7 \times 1} \\
 0_{1 \times 7} & 512 z_c^2 L_c^2 & 0 & 3072 z_c^2 L_c - 8192 z_c^2 L_c^2 \\
 0_{1 \times 7} & 0 & 0 & 0 \\
 0_{1 \times 7} & 3072 z_c^2 L_c - 8192 z_c^2 L_c^2 & 0 & 131072 z_c^2 L_c^2 - 98304 z_c^2 L_c + 18432 z_c^2 \\
\end{array}
\right) \, , \\[8pt]
\hat{G}_{dg}^{bc} &=& \hat{G}_{dg}^{cb \, \dagger} = \left(
\begin{array}{cccccccccc}
 0_{1 \times 7} & 128 z_c L_c & 0 & 256 z_c (3 - 8 L_c) \\
 0_{1 \times 7} & 2560 z_c L_c & 0 & -5120 z_c (8 L_c - 3) \\
 0_{1 \times 7} & -128 z_c L_b L_c & 0 & 256 z_c L_b (8 L_c - 3) \\
 0_{1 \times 7} & 128 z_c (4 L_b - 1) L_c & 0 & -256 z_c (4 L_b - 1) (8 L_c - 3) \\
 0_{1 \times 7} & 2048 z_c (5 - 8 L_b) L_c & 0 & 4096 z_c (8 L_b - 5) (8 L_c - 3) \\
\end{array} \right) \, , \; \; \hat{G}_{dg}^{GG} \,=\, \frac{32}{\tilde{\alpha}_s^2}\, ,\\[8pt]
\hat{G}_{dg}^{bb} &=& \left(
\begin{array}{ccccc}
 32 & 640 & 0 & -32 & 2560 \\
 640 & 12800 & 0 & -640 & 51200 \\
 0 & 0 & 0 & 0 & 0 \\
 -32 & -640 & 0 & 32 & -2560 \\
 2560 & 51200 & 0 & -2560 & 204800 \\
\end{array}
\right) + L_b \left(
\begin{array}{ccccc}
 0 & 0 & -32 & 128 & -4096 \\
 0 & 0 & -640 & 2560 & -81920 \\
 -32 & -640 & 0 & 32 & -2560 \\
 128 & 2560 & 32 & -256 & 14336 \\
 -4096 & -81920 & -2560 & 14336 & -655360 \\
\end{array}
\right) \notag \\[8pt]
& & + L_b^2 \left(
\begin{array}{ccccc}
 0 & 0 & 0 & 0 & 0 \\
 0 & 0 & 0 & 0 & 0 \\
 0 & 0 & 32 & -128 & 4096 \\
 0 & 0 & -128 & 512 & -16384 \\
 0 & 0 & 4096 & -16384 & 524288 \\
\end{array}
\right)
\, , \\[8pt]
\hat{G}_{dg}^{Gc} &=& \hat{G}_{dg}^{cG} =  \frac{1}{\tilde{\alpha}_s} \left( 
\begin{array}{ccccc}
0_{1 \times 6} & 0 & 128 z_c L_c & 0 & 768 z_c -2048 z_c L_c
\end{array} \right) \, , \\[8pt]
\hat{G}_{dg}^{Gb} &=& \hat{G}_{dg}^{bG} =  \frac{1}{\tilde{\alpha}_s} \left( 
\begin{array}{ccccc}
32 & 640 & -32 L_b & 128 L_b -32 & 2560 -4096 L_b
\end{array} \right) \, .
}

\section{Phenomenological Impact}
\label{sec:applications}

Having computed all NLO contributions to the partonic $b$-quark decay rate from light-particle final states, we now assess their potential phenomenological implications. For that matter, we study how these NLO corrections affect the theoretical sensitivity of the lifetime to the different directions in the parameter space of Wilson coefficients. 
In particular, decays involving unflavored final states ($b\to d \bar q q$ and $b\to d g$) are of special interest, given that they involve one single large set of Wilson coefficients (also mixed by RG evolution), $\big\{\mathcal{C}^{dbX}_{i}\big\}$, which is also involved in other flavor observables of remarkable interest such as semi-leptonic $B$-meson decays. We then combine all of these contributions in a single~$41\times41$ matrix:
\begin{equation}
\label{eq:decay_rate_dbX}
    \Gamma(b \to d \bar u u) + \Gamma(b \to d \bar s s) + \Gamma(b \to d \bar d d) + \Gamma(b \to d g) = \sum_{i,j} (\mathcal{C}_i^{dbX\dagger} \mathcal{C}^{dbX}_j + \mathcal{C}_{i'}^{dbX\dagger} \mathcal{C}^{dbX}_{j'}) \, \mathcal{G}_{ij}[d \bar qq] \, ,
\end{equation}
where the combined matrix reads:
\begin{equation}
    \hat{\mathcal{G}}[d \bar qq] \equiv \Gamma_0 \left( \hat{G}_{CC}[d \bar qq]
    + \hat{G}_P[d \bar uu] + \hat{G}_P[d \bar ss] + \hat{G}_P[d \bar dd] + \hat{G}_P[dg]\right) \, ,
\end{equation}
and we have combined all current-current contributions into a large matrix, as anticipated in the previous section:
\begin{equation}
\hat{G}_{CC}[d \bar qq] \equiv 
\left(
    \begin{array}{ccccc}
    \hat{G}_{CC}[d \bar uu] & 0_{10\times10} & 0_{10\times10} & 0_{10\times5} & 0_{10\times6} \\
    0_{10\times10} & \hat{G}_{CC}[d \bar ss] & 0_{10\times10} & 0_{10\times5} & 0_{10\times6} \\
    0_{10\times10} & 0_{10\times10} & 0_{10\times10} & 0_{10\times5} & 0_{10\times6} \\
    0_{5\times10} & 0_{5\times10} & 0_{5\times10} & \hat{G}_{CC}[d \bar dd] & 0_{5\times6} \\
    0_{6\times10} & 0_{6\times10} & 0_{6\times10} & 0_{6\times5} & 0_{6\times 6} 
    \end{array}
    \right) \, .
\end{equation}
Of course, the matrix $\hat{\mathcal{G}}[d \bar qq]$ can also be perturbatively expanded in terms of $\tilde\alpha_s$, with coefficients $\hat{\mathcal{G}}^{(k)}[d \bar qq]$.
In order to provide constraints on the Wilson coefficients, we can compare this expression to the experimentally measured non-leptonic decay width of a $B$-meson, which can be defined as the full decay width minus all contributions from semi-leptonic and rare channels. The rare channels, however, only contribute a small fraction, so we do not consider them in our analysis. Similar to~Ref.~\cite{Meiser:2024zea}, we use the experimental non-leptonic decay width\footnote{Since we only consider the leading power term in the HQE, our bound in principle is valid for any $B$-meson. However, we found the decay width of the $B^-$ to be the lowest and thus the most constraining. In any case, our bounds do not depend strongly on the choice of the $B$-meson considered.}
\begin{align}
    \Gamma_{\text{exp}}^{\text{NL}}(B^-) \approx \Gamma(B^-)_{\text{PDG}} - \Gamma(B^- \to X \ell^- \nu)_{\text{PDG}} = 3.035\times 10^{-13} \, \text{GeV} \, , 
\end{align}
where in the semi-leptonic rate we consider the sum of all three lepton flavors in the final state.
The uncertainty associated with these measurements is much smaller than the theoretical renormalization-scale uncertainty carried by our calculation, with which we neglect it henceforth. 
Since we only include a subset of all possible non-leptonic channels, we demand
\begin{align}
    \sum_{i,j} (\mathcal{C}_i^{dbX\dagger} \mathcal{C}^{dbX}_j + \mathcal{C}_{i'}^{dbX\dagger} \mathcal{C}^{dbX}_{j'}) \, \mathcal{G}_{ij}[d \bar qq] \leq \Gamma_{\text{exp}}^{\text{NL}}(B^-) \, . \label{eq:defEllipsoid}
\end{align}
That is, the sum of the decay widths of the charmless $b \to d \bar q q (dg)$ channels can never exceed the full non-leptonic decay width. Of course, in this form the bound will be very loose since we do not consider all possible channels. However, the point is not to get a very tight bound on the Wilson coefficients but to have as many constrained directions as possible, even if just loosely constrained.
It is also worth mentioning that including more channels on the LHS, e.g. $b \to s \bar q q (sg)$, $b \to d \bar s d$, $b \to s \bar d s$, $b \to d(s)\bar u c$ or $b \to d(s)\bar c c$, would lead to higher-dimensional but tighter constraints, up to the point of saturating the inequality upon combination of all inclusive channels. At the same time, an equation analogous to~Eq.~(\ref{eq:defEllipsoid}) would also apply for each of them separately.

In the following, we will analyze the implications of~Eq.~(\ref{eq:defEllipsoid}) by studying the spectrum of eigenvalues of the matrix $\hat{\mathcal{G}}[d\bar{q}q]$, and presenting the two-dimensional projections of the open ellipsoid it defines. 
We then discuss the potential of including massive cuts in the analysis. 
Finally, we briefly comment on the analogous matrix $\hat{\mathcal{G}}[d\bar{s}d]$. 

\subsection{Constraints on $\big\{\mathcal{C}_i^{dbX}\big\}$}
As a first step in our analysis, we diagonalize the matrix $\hat{\mathcal{G}}[d\bar{q}q]$ at the numerical level to study its eigenvalues ($\lambda_i$), and compare them before and after including the NLO corrections. Positive eigenvalues will then correspond to (orthogonal) directions where the lifetime provides a constraint, while vanishing $-$or negative, as we will later discuss$-$ eigenvalues will signal unconstrained directions. 

\begin{figure}[t]
    \centering
    \includegraphics[width=\textwidth]{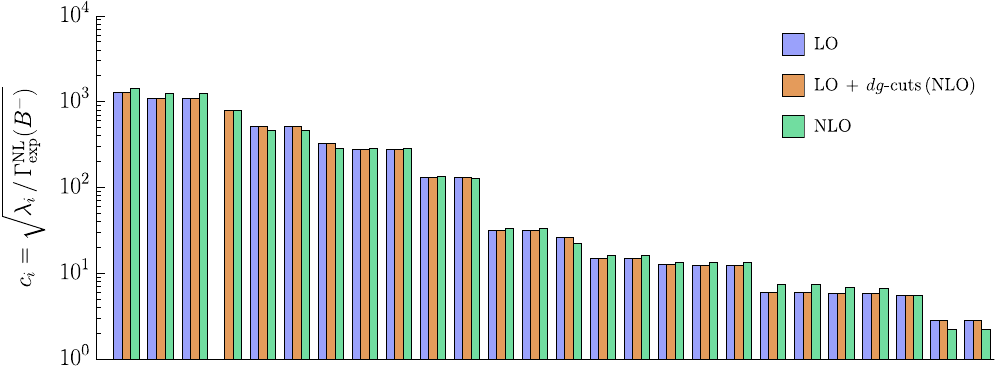}
    \caption{Spectrum of eigenvalues of $\hat{\mathcal{G}}[d \bar qq]$ for the case of the LO matrix, the matrix including also the contributions from two-particle cuts, and the matrix including all NLO contributions from charmless final states. The constraint along the direction of $\mathcal{C}_\star$ is introduced solely by $\hat{\mathcal{G}}[d g]$. We normalize the eigenvalues with $\Gamma_{\rm exp}^{\rm NL}(B^-)$ and take the square root to obtain the inverse of the associated maximal constraints, $c_i$.}
    \label{fig:eigenplot_dL_All}
\end{figure}
\begin{figure}[t]
    \centering
    \includegraphics[width=0.95\textwidth]{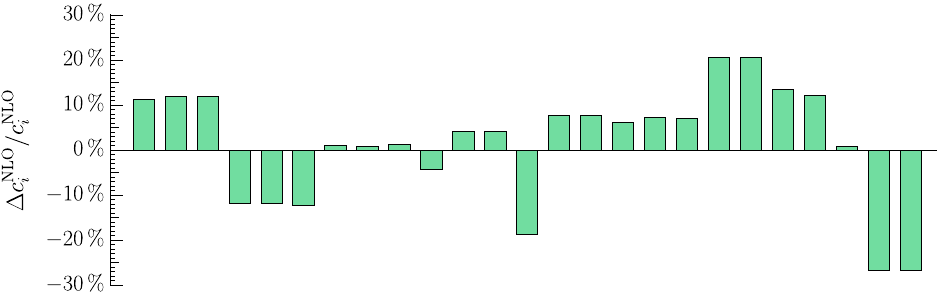}
    \caption{Relative corrections to the constraint put by each eigenvalue due to the NLO calculation, $\Delta c_i^{\rm NLO}/c_i^{\rm NLO} = 1 - c_i^{\rm LO}/c_i^{\rm NLO} $, excluding the bound on $\mathcal{C}_\star$, which appears only at NLO. We display them ordered by decreasing magnitude, such that each position in the horizontal axis matches the corresponding position in~Fig.~\ref{fig:eigenplot_dL_All}. }
    \label{fig:eigenplot_dL_Rel}
\end{figure}

For the numerical analysis, we use the $\overline{\rm MS}$ values of $\alpha_s$, $m_b$ and $m_c$ at $\mu = 4$ GeV, obtained using \texttt{RunDec}~\cite{Chetyrkin:2000yt,Herren:2017osy}, which yield:
\begin{equation}
    \alpha_s(4\,{\rm GeV}) = 0.229 \,, \quad m_b(4\,{\rm GeV}) = 4.198 \,{\rm GeV} \,, \quad m_c(4\,{\rm GeV}) = 0.939 \,{\rm GeV} \,.
\end{equation}
With this setup, out of the 41 available Wilson coefficients, the LO matrix defines 25 non-flat directions. The two-particle cuts introduce a single strong constraint along a new orthogonal direction, which we label as $\lambda_\star$, while leaving all other constraints largely the same. This direction corresponds analytically, up to normalization, to
\begin{align}
    \mathcal{C}_\star = \mathcal{C}_{8g}^{d} &\,+ \tilde{\alpha}_s \bigg[ \big( 80 - 128L_b\big)\,\mathcal{C}_{9}^{dbbb} - \big( 1 - 4L_b\big)\,\mathcal{C}_{7}^{dbbb} - L_b\,\mathcal{C}_{5}^{dbbb} + 20\,\mathcal{C}_{3}^{dbbb} + \mathcal{C}_{1}^{dbbb}\bigg] \notag \\
    &\,+ \tilde{\alpha}_s z_c \bigg[ \big( 24 - 64L_c\big)\,\mathcal{C}_{10}^{dbcc} + 4 L_c\,\mathcal{C}_{8}^{dbcc}\bigg]\, ,
\label{eq:Cstar}
\end{align}
which concerns mostly $\big\{ \mathcal{C}_{8g}^{d} , \mathcal{C}_{9}^{dbbb} , \mathcal{C}_{3}^{dbbb}\big\}$ up to percent-level effects. Meanwhile, the (four- and) three-particle one-loop contributions provide sizable corrections to the LO bounds and to the corresponding directions. Let us note that many of the LO eigenvalues are of the order of a few-percent or smaller with respect to the biggest ones, which might have rendered them unstable against NLO corrections. However, we report a stabilization of all LO constraints in this sense, up to NNLO corrections. The resulting eigenvalue spectrum at NLO features 26 non-vanishing eigenvalues, constraining all $dbuu$, $dbss$ and $dbdd$ Wilson coefficients, and $\mathcal{C}_\star$. This is shown in~Fig.~\ref{fig:eigenplot_dL_All} in terms of their constraining power,
\begin{equation}
    c_i = \sqrt{\frac{\lambda_i}{\Gamma_{\rm exp}^{\rm NL}(B^-)}} \, ,
\end{equation}
such that the bounds put on each direction are given by $1/c_i$. 
In~Fig.~\ref{fig:eigenplot_dL_Rel} we display the relative differences between the LO and NLO eigenvalues (again, in terms of $c_i$), which offer better visualization of the corrections induced in the respective bounds.

Additionally to the direction $\mathcal{C_\star}$, that was introduced by the two particle cuts, also the three-particle NLO corrections yield several additional non-flat directions, involving $dbcc$ and $dbbb$ coefficients, many of these exhibit negative eigenvalues.
The reason behind this is that most directions are mainly constrained by the $CC$ insertions; that is, the Wilson coefficient of an operator with flavor content $dbqq$ is mainly constrained by the $CC$ diagrams with $d \bar qq$ cuts. 
Given that we include neither $d \bar c c$ nor $d \bar b b$ cuts (the latter are kinematically forbidden), the corresponding directions in $\hat{G}[d \bar q q]$ receive only sub-leading bounds due to cross-flavor penguin insertions. 
As such, these eigenvalues would change drastically upon including $d \bar c c$ cuts and other contributions constraining predominantly the set of $dbbb$ operators (or the dipole $\mathcal{C}_{8g}^d$), as could be the case for the NNLO $dgg$ cuts. 
For this reason, we consider all such directions unstable, and we have manually put the corresponding Wilson coefficients (i.e. all $dbcc$ and $dbbb$ except $\mathcal{C}_\star$) to zero for our analysis, effectively removing their rows and columns from $\mathcal{G}[d \bar q q]$. 

\begin{figure}[t]
    \centering
    \includegraphics[width=\textwidth]{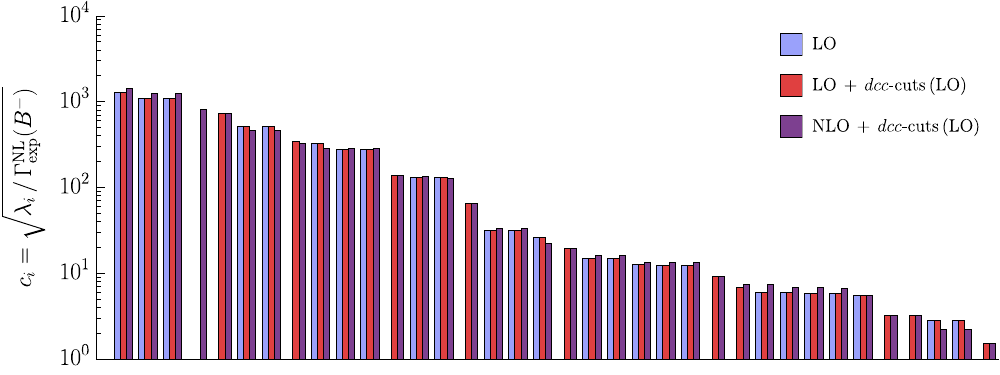}
    \caption{Spectrum of eigenvalues of $\hat{\mathcal{G}}[d \bar qq]$ upon inclusion of the LO contributions from $d \bar c c$ final states. We normalize the eigenvalues with $\Gamma_{\rm exp}^{\rm NL}(B^-)$ and take the square root to obtain the inverse of the associated maximal constraints, $c_i$. Note that the 10 eigenvalues introduced by the $d\bar c c$ final-state contributions, covering all $dbcc$ directions, are barely affected by our NLO corrections.}
    \label{fig:eigenplot_dL+dcc_All}
\end{figure}

As explicit proof of the unstable nature of the constraints on $dbcc$ operators obtained purely from $\hat{\mathcal{G}}[d\bar q q]$, we have also computed the LO contribution to the decay rate due to $d \bar c c$ final states, and combined it with $\hat{\mathcal{G}}[d\bar q q]$ to study the corresponding eigenvalue spectrum. 
We show this spectrum in~Fig.~\ref{fig:eigenplot_dL+dcc_All}, where only the five unstable directions among $\mathcal{C}_{i}^{dbbb}$ and $\mathcal{C}_{8g}^d$ are put to zero. What we find, as anticipated in the previous discussion, is that the 10 new constraints that emerge in this case, corresponding essentially to the $dbcc$ operators, are clearly dominated by the $d\bar c c$ cuts and show only minor changes after combination with the one-loop three-particle corrections.

\begin{figure}[t]
    \centering
    \includegraphics[width=\textwidth]{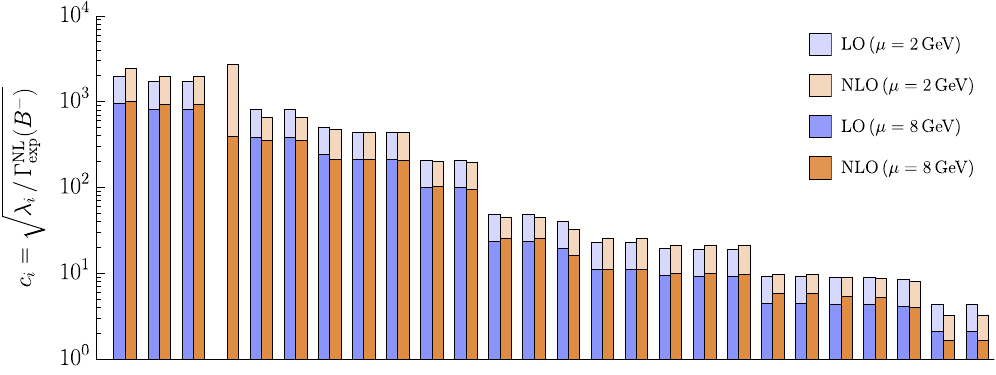}
    \caption{Spectrum of eigenvalues of $\hat{\mathcal{G}}[d \bar qq]$ for the case of the LO matrix and the matrix including all NLO contributions from light-particle final states, for $\mu=\{\mu_0/2,2\mu_0\}$, with $\mu_0=4$ GeV being our reference central value. We normalize the eigenvalues with $\Gamma_{\rm exp}^{\rm NL}(b)$ and take the square root to obtain the inverse of the associated maximal constraints, $c_i$.}
    \label{fig:eigenplot_dL_scales_All}
\end{figure}

Concerning the dependence on the $\overline{\rm MS}$ renormalization scale $\mu$ of our results, and the comparison between the LO and NLO behavior, in~Fig.~\ref{fig:eigenplot_dL_scales_All} we show the results of a simplified analysis applied to the case of $\hat{\mathcal{G}}[d \bar qq]$, varying $\mu$ from $2$ GeV to $8$ GeV. We do not consider, however, the use of better-behaved schemes for the definition of the quark masses $m_b$ and $m_c$~\cite{Egner:2024lay}, nor the RG evolution of the Wilson coefficients. 
In terms of the scale dependence of the NLO decay rate with respect to the LO, this simple analysis reveals no clear pattern of improvement, with about half of the bounds (mostly the smaller eigenvalues) being more stable at NLO, and the rest acquiring a larger scale dependence. Furthermore, the $\mu$-dependence of $\lambda_\star$ seems to be much higher than that of any other eigenvalue. Still, we refrain from making any definitive claims, given that a better choice of scheme for $m_b$, the inclusion of the RG effects due to the Wilson coefficients, and even the inclusion of the contributions from massive final states ($b \to d \bar c c$), can greatly modify this behavior. Instead, our intention with~Fig.~\ref{fig:eigenplot_dL_scales_All} is merely to show that, while the individual eigenvalues might change significantly, the qualitative picture defined by~Fig.~\ref{fig:eigenplot_dL_All} is rather persistent, and not strongly dependent on the choice of $\mu$.

\begin{figure}[t]
    \includegraphics[width=\textwidth]{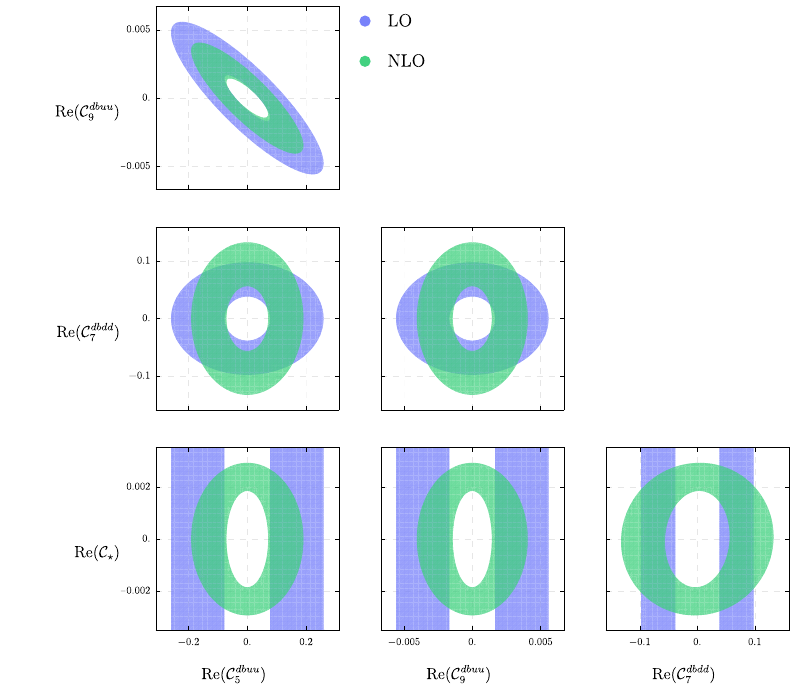}
    \caption{Two-dimensional projections of the open ellipsoid defined by~Eq.~\ref{eq:defEllipsoid} for a small subset of the parameter space. The colored regions visualize the impact of the scale variation for $\mu \in [2 \, \text{GeV}, \, 8 \, \text{GeV}]$. A thin band means small scale variation while a wide band indicates large scale variation. Note that here we set $\mathcal{C}_i^{dbcc} = \mathcal{C}_i^{dbbb} = 0$ for all $i$.}
    \label{fig:triangle}
\end{figure}
In addition to the discussion on the spectrum, we study the two-dimensional projections of the high-dimensional ellipsoid defined by~Eq.~\ref{eq:defEllipsoid}, shown in~Fig.~\ref{fig:triangle}, with the region inside the ellipsoid being allowed parameter space and the region outside being excluded. Here we use the full charmless LO and NLO results, again setting the Wilson coefficients of the unconstrained $dbcc$ and $dbbb$ operators to zero, and illustrate their respective scale variations. For different values of $\mu$, the ellipsoids change their shape and size and hence varying the renormalization scale within $\mu \in [2 \, \text{GeV}, \, 8 \, \text{GeV}]$ generates a family of ellipsoids, which is illustrated with the colored regions. The inner boundaries of the colored regions correspond to $\mu = 2 \, \text{GeV}$, while the outer boundaries correspond to $\mu = 8 \, \text{GeV}$. 
In line with the findings shown in~Fig.~\ref{fig:eigenplot_dL_scales_All}, the magnitude of the bounds at LO and NLO is rather strongly dependent on the renormalization scale and no systematic improvement is seen when including NLO corrections. 
Nevertheless, we do see that some of the constraints (e.g. $\mathcal{C}_5^{dbuu}$ - $\mathcal{C}_7^{dbdd}$ plane) change their shape. This happens because one of the bounds gets tighter, while the other gets looser upon including NLO corrections. This is in line with the findings illustrated in~Fig.~\ref{fig:eigenplot_dL_Rel} that the NLO corrections push some of the eigenvalues up and some others down.
The most notable change, however, is that the direction $\mathcal{C_\star}$, defined in~Eq.~\ref{eq:Cstar}, is completely unconstrained at LO but is quite well constrained at NLO. 
We point out again that while relative NLO corrections of up to $\sim 25 \, \%$ do constitute large NLO corrections, they still do not significantly change the qualitative picture. 
The point of our constraint is not to give very stringent bounds on the Wilson coefficients but rather to give some sort of loose bound on as many directions as possible such that in a BSM fit one could use our bound as a prior on the Wilson coefficients with finite parameter space volume. 
This essentially cuts off directions unconstrained by other observables.
Hence, the exact size of the bounds is not as important as the number of constrained directions and their rough order of magnitude.

Let us reiterate, that in our convention the CKM matrix elements are absorbed into the definitions of the Wilson coefficients, which might mislead one to interpret the bounds that we give as much stronger than they really are. This is especially true for Wilson coefficients that would usually come with a strong CKM supression.

\subsection{Constraints on $\big\{\mathcal{C}_i^{dbds}\big\}$}

We carry out a similar spectral analysis for the matrix $\hat{\mathcal{G}}[d\bar s d] \equiv \Gamma_0 \, \hat{G}_{CC}[d\bar s d]$, which reveals 5 different eigenvalues already at LO; that is, it provides constraints for all $dbds$ directions. The one-loop corrections come only from current-current diagrams in this case, and they provide small but relevant corrections to the eigenvalues (of up to $25\%$). Still, they leave the overall qualitative picture unchanged, which again increases the confidence on the corresponding bounds, in spite of the fact that they span up to three orders of magnitude, and suggest that no further qualitative changes are to be expected from a NNLO calculation.

\section{Outlook and Conclusions}
\label{sec:conclusions}

The non-leptonic $B$-meson lifetime receives contributions from all 40 Wilson coefficients in the $dbcu$ and $sbcu$ sectors of the Weak Effective Theory, and has been shown to be a powerful constraint complementary to exclusive non-leptonic observables of the type $b \to c \bar{u}d$ and $b \to c \bar{u}s$. Motivated by this fact, we have computed the full set of NLO QCD corrections to the inclusive charmless non-leptonic lifetime at leading power in the Heavy Quark Expansion.

We have shown how our results constrain the Wilson coefficients of operators that are also relevant for semi-leptonic decays, and found that, contrary to the $d(s)bcu$ case, not all directions are constrained. 
We find in particular that the LO contributions constrain 25 out of 41 directions, and that the NLO contributions introduce one additional non-flat direction, while leaving all bounds present at LO qualitatively the same. 
This suggests that the corresponding bounds are stable under NNLO corrections or higher, thus validating the approach taken in~Ref.~\cite{Meiser:2024zea} with respect to the use of inclusive lifetimes as complementary constraints.

We have also discussed the importance of the  $d\bar{c}c$ cuts at LO. 
However, in order to be consistent and to ensure the stabilization of the 10 additional bounds, it would be highly desirable to compute also the NLO corrections corresponding to these cuts. 
A first step could be the inclusion of the NLO corrections within the limit of $m_c \to 0$ while keeping the charm mass in the LO result. This can be extracted directly from our results.
Additionally, including NNLO corrections in the form of $dgg$-cuts or other types of final-state contributions might also prove useful in constraining the remaining five flat directions among the $dbbb$ and dipole operators. 
However, such contributions would need to be accompanied by the corresponding NNLO corrections to the final states we consider here, both for consistency in the precision and for proper renormalization of the results.

Another important improvement of our bounds could be achieved by considering new physics in a specific sub-sector of the LEFT/WET, which allows for the formulation of~Eq.(\ref{eq:defEllipsoid}) only in terms of the channels affected by NP, thereby making the bound much tighter. At the practical level, this can be achieved by subtracting the SM predictions for all (inclusive) channels from $\Gamma_{\rm exp}^{\rm NL}$ where NP does not enter.

Finally, we point out that for the purpose of this computation it was convenient to work in the Bern basis. 
However, integrating our results into the full NLO and NNLL LEFT/SMEFT pipeline requires presenting the results in the JMS basis of~Ref.~\cite{Jenkins:2017jig}. 
This will be part of a forthcoming publication~\cite{Meiser:2026}.

\acknowledgments

This work is funded by the Spanish
MCIN/AEI/10.13039/501100011033 through the following grants: 
grant CNS2022-135262 funded by the “European Union NextGenerationEU/PRTR”,
grant CEX2019-000918-M through the “Unit of Excellence Mar\'ia de Maeztu 2020-2023” award to the Institute of Cosmos Sciences,
and grant PID2022-136224NB-C21.
S.M. is supported by the Studienstiftung des deutschen Volkes. 
P.M. is supported by the Spanish MCIN/AEI/10.13039/501100011033 through grant PRE2022-103999.
A.P. is supported by a FPU fellowship from the Spanish Ministry of Science, Innovation and Universities (FPU23/02972).

\appendix

\section{Cutkosky Rules}
\label{app:Cutkosky}

In order to evaluate the contribution of a Feynman diagram (renormalized in the $\overline{\rm MS}$ scheme) to the imaginary part of an $A \to B$ scattering amplitude, we proceed through the following steps:
\begin{enumerate}
    \item Consider all possible ways of splitting the diagram in two by cutting internal propagators. A valid cut must slice only through propagators of stable particles (i.e. with vanishing decay width up to the relevant level of precision).
    \item Retain only the cuts that divide the diagram into two amputated diagrams, which we refer to as ``amputated cuts''.
    \item Apply the usual Feynman rules on every cut diagram, barring the change in propagator prescription ${+i\epsilon \to -i\epsilon}$ to the right of the cut.
    \item Before loop integration, for each cut propagator of a field $\psi$ with momentum $p$, mass parameter $m$ and pole mass $M$, apply the following replacement: 
    \eq{\frac{1}{p^2 - m^2 + i\epsilon} \rightarrow -2\pi i \delta(p^2-M^2)\theta(p^0)\frac{Z_\psi^{\rm OS}}{Z_\psi^{\overline{\rm MS}}},}
    where $Z_\psi^{R}$ is the field strength renormalization constant of $\psi$ in scheme $R$.
    \item Evaluate all loop integrals to obtain the contribution of each cut to the discontinuity across the corresponding branch cut of the amplitude: \eq{\text{Disc}\,i\mathcal{M}_{A\to B} = -2\,\text{Im}\,\mathcal{M}_{A\to B}.}
    \item Sum over all valid cuts to obtain the final result.
\end{enumerate}
Let us recall that the pole mass and the Lagrangian mass are related by $M = m + \Sigma^{\overline{\rm MS}}_\psi(M)$ in case of fermions, and by $M = m + \Pi^{\overline{\rm MS}}_\psi(M^2)$ for bosons; where $\Sigma^{\overline{\rm MS}}_\psi(\slashed{p})$ and $\Pi^{\overline{\rm MS}}_\psi(p^2)$ are the fermion and boson $\overline{\rm MS}$-renormalized self-energies (SE), respectively. These rules apply all the same to any renormalization scheme $R$, and one may write $\overline{\rm MS}\to R$.

Note that the second and fourth rules are apparently different from the original Cutkosky rules~\cite{Cutkosky:1960sp}, for which non-amputated cuts must also be included, and the replacement rule reads instead:
\eq{
\label{eq:Cutkosky_Old}
\frac{1}{p^2 - m^2 + i\epsilon} \rightarrow -2\pi i \delta(p^2-m^2)\theta(p^0). 
}
It can be proven that both sets of rules are completely equivalent, and that the factor ${Z_\psi^{\rm OS}}/{Z_\psi^{\overline{\rm MS}}}$ in the fourth rule accounts exactly for the resummation of all cuts dismissed by the second rule. These are cut diagrams where a dressed propagator is sliced through a single line, leading to a non-amputated cut. To prove this equivalence, we first define the shorthands:
\eq{
S(\slashed{p}) \equiv \frac{(\slashed{p}+m)}{p^2-m^2+i\epsilon} \,, \qquad \Delta(\slashed{p}) \equiv 2\pi\delta(p^2-m^2)\theta(p^0)(\slashed{p}+m)\,,
}
and let $\mathcal{K}_{\rm 1p}$ represent the sum over all possible applications of the replacement~\ref{eq:Cutkosky_Old} to a single propagator in a certain self-energy chain. All non-amputated cuts dismissed by our second rule are proportional to an application of $\mathcal{K}_{\rm 1p}$ over a loop-corrected propagator. Let us now study the overall result of applying such one-particle cuts to a fermion line containing a single SE insertion:
\eqa{
\mathcal{K}_{\rm 1p}\big[iS(\slashed{p}) (-i)\Sigma^{\overline{\rm MS}}_\psi(\slashed{p}) iS(\slashed{p})\big] & = & \Delta(\slashed{p}) \Sigma^{\overline{\rm MS}}_\psi(\slashed{p}) S(\slashed{p})^* + S(\slashed{p}) \Sigma^{\overline{\rm MS}}_\psi(\slashed{p})\Delta(\slashed{p}) \notag \\*
& = & i(S(\slashed{p})-S(\slashed{p})^*) \Sigma^{\overline{\rm MS}}_\psi(\slashed{p}) S(\slashed{p})^* + S(\slashed{p}) \Sigma^{\overline{\rm MS}}_\psi(\slashed{p})i(S(\slashed{p})-S(\slashed{p})^*) \notag \\*
& = & i  \Sigma^{\overline{\rm MS}}_\psi(\slashed{p}) \Big[S(\slashed{p})^2 - \big(S(\slashed{p})^*\big)^2\Big] \notag \\*
& = & -i  \Sigma^{\overline{\rm MS}}_\psi(\slashed{p}) \frac{\partial}{\partial \slashed{p}}\Big[S(\slashed{p}) - S(\slashed{p})^*\Big] \notag \\*
& = & -\Sigma^{\overline{\rm MS}}_\psi(\slashed{p}) \frac{\partial}{\partial \slashed{p}}\Delta(\slashed{p}) \,,
}
where we have used the fact that the SE is real at the pole, and that it commutes with $\slashed{p}$.
An analogous strategy leads to showing that, in case of having $n$ insertions of the SE in the same fermion line,
\eq{
\mathcal{K}_{\rm 1p}\big[iS^{(n)}(\slashed{p})\big] \equiv \mathcal{K}_{\rm 1p}\big[iS(\slashed{p}) \underbrace{(-i)\Sigma^{\overline{\rm MS}}_\psi(\slashed{p}) iS(\slashed{p}) \cdots (-i)\Sigma^{\overline{\rm MS}}_\psi(\slashed{p}) iS(\slashed{p})}_{n}\big] = \Big[-\Sigma^{\overline{\rm MS}}_\psi(\slashed{p})\Big]^n \frac{1}{n!}\frac{\partial^n}{\partial \slashed{p}^n}\Delta(\slashed{p}) \,.
}
Resumming all possible configurations of one-particle cuts over all possibilities of SE insertions, one obtains:
\eqa{
\sum_{n=0}^\infty \mathcal{K}_{\rm 1p}\big[iS^{(n)}(\slashed{p})\big] & = & 2\pi \delta\Big(\slashed{p} - m - \Sigma^{\overline{\rm MS}}_\psi(\slashed{p})\Big) \theta(p^0) \notag \\*
& = & 2\pi \delta(p^2-M^2) \theta(p^0) (\slashed{p}+M) \bigg[1 - \frac{\partial}{\partial \slashed{p}}\Sigma^{\overline{\rm MS}}_\psi(\slashed{p}) \Big|_{\slashed{p}=M}\bigg]^{-1} \notag \\*
& = & \mathcal{K}_{\rm 1p} \bigg[\frac{i}{\slashed{p}-M+i\epsilon}\bigg] \,\frac{Z_\psi^{\rm OS}}{Z_\psi^{\overline{\rm MS}}} \,, \label{eq:Cutkosky_resummation}
}
where the Dirac delta in the first line is formally using $\slashed{p}$ as a variable, but is only evaluated through its form in the second line.
The proof of the first equality, corresponding to the evaluation of the resummed series, follows from applying the LHS onto a test function $\phi(\slashed{p})$ within some loop integral (implicit),
\eqa{
\sum_{n=0}^\infty \frac{1}{n!} \Big[-\Sigma^{\overline{\rm MS}}_\psi(\slashed{p})\Big]^n \, \bigg[\frac{\partial^n}{\partial \slashed{p}^n}\Delta(\slashed{p}) \bigg] \phi(\slashed{p}) & = & \sum_{n=0}^\infty \frac{2\pi}{n!} \frac{\partial^n}{\partial \slashed{p}^n} \bigg(\Big[\Sigma^{\overline{\rm MS}}_\psi(\slashed{p})\Big]^n \phi(\slashed{p})\bigg) \bigg|_{\slashed{p}=m}
\notag \\*
& = & -i \sum_{n=0}^\infty \oint_C  d\xi \frac{1}{(\xi-m)^{n+1}} \Big[\Sigma^{\overline{\rm MS}}_\psi(\xi)\Big]^n \phi(\xi)
\notag \\*
& = & -i \oint_C  d\xi \frac{1}{(\xi-m)} \sum_{n=0}^\infty \Big[\frac{1}{(\xi-m)}\Sigma^{\overline{\rm MS}}_\psi(\xi)\Big]^n \phi(\xi)
\notag \\*
& = & -i \oint_C d\xi \frac{1}{\xi -m - \Sigma^{\overline{\rm MS}}_\psi(\xi)} \phi(\xi) 
\notag \\*
& = & 2\pi\bigg[1 - \frac{\partial}{\partial \slashed{p}}\Sigma^{\overline{\rm MS}}_\psi(\slashed{p}) \Big|_{\slashed{p}=M}\bigg]^{-1} \phi(M) \;,
}
which is the result of applying the delta function of Eq.~(\ref{eq:Cutkosky_resummation}) to the test function.
This derivation requires that the contour $C$ encompasses the pole at the physical mass, which must be isolated. This does not happen in the presence of massless fields, in which case an IR regulator must be introduced\footnote{For instance, a mass regulator $\lambda$ for the massless fields would allow already for the definition of a contour around the simple pole, taking the limit $\lambda \rightarrow 0$ only at the level of the final amplitudes. Working in $d<4$ within dimensional regularization, as we do in the main text, achieves the same goal.}.

This proves that, for a diagram containing a fermion propagator, the overall contribution due to all cuts effecting a $\mathcal{K}_{\rm 1p}$ on it, after including all possible loop-corrections, factorizes into the contribution of our fourth rule. A completely analogous proof follows for the case of boson SE insertions sliced only through one-particle cuts. It then follows that if one uses the modified replacement rule, all non-amputated cuts are automatically taken into account, and need not be directly included.

The previous derivation fails, however, if the SE has a non-vanishing imaginary part at the pole. In such a case, however, the corresponding particle is unstable, and thus all valid cuts will have to avoid its internal propagators anyway. The two sets of Cutkosky rules remain thus equivalent also in this case.

\section{Details of the Loop Integration}
\label{app:Loops}
In our calculation, we encounter a number of Passarino-Veltman integrals. Our convention for the $B_0$ and $C_0$ integrals is
\begin{align}
  B_0(p_1^2, m_0^2, m_1^2) &= \tilde{\mu}^{4-d} \int \frac{d^4k}{(2\pi)^d} \frac{1}{(k^2-m_0^2) [(k+p_1)^2 - m_1^2]} \, , \\
  C_0(p_1^2,(p_1-p_2)^2, p_2^2, m_0^2, m_1^2,m_2^2) &=  \tilde{\mu}^{4-d} \int \frac{d^4k}{(2\pi)^d} \frac{1}{(k^2-m_0^2) [(k+p_1)^2 - m_1^2] [(k+p_2)^2 - m_2^2]} \, ,
\end{align}
with $\tilde{\mu} = \mu \sqrt{e^{\gamma_E}/(4\pi)}$. The expressions evaluated for the parameter values that appear in our computation are
\eqa{
    B_0(s,0,0) &=& \frac{i}{16 \pi^2}\left(\frac{4\pi \tilde{\mu}^2}{-s} \right)^{2-d/2}\Gamma(2-d/2) \beta(d/2-1,d/2-1)\, , \\
     B_0(0,m^2,m^2) &=& \frac{i}{16 \pi^2} \left(\frac{4 \pi \tilde{\mu}^2}{m^2}\right)^{2-d/2} \Gamma(2-d/2) \, , \\
    B_0(s, 0, m^2) &=& \frac{i}{16 \pi^2} \left (\frac{4\pi \tilde{\mu}^2}{m^2}\right)^{2-d/2} \Gamma(2-d/2) \beta(1,d/2-1) \, {}_2F_1(1,2-d/2;d/2;s/m^2) \, , \\
    B_0(s, m^2, m^2) &=& \frac{i}{16 \pi^2}  \left(\frac{4 \pi \tilde{\mu}^2}{m^2}\right)^{2-\frac{d}{2}} \, \Gamma \left(2-\frac{d}{2}\right) {}_2F_1\left(1,2-\frac{d}{2};\frac{3}{2};\frac{s}{4 m^2}\right) \, , \\
    C_0(0,0,s,0,0,0) &=& \frac{i}{16 \pi^2} \frac{1}{s(d-4)} \left(\frac{4 \pi \tilde{\mu}^2}{-s}\right)^{2-d/2} \Gamma(3-d/2) \beta(d/2-2,d/2-2) \, , \\
    C_0(m^2,0,s,m^2,0,0) &=& - \frac{i}{16\pi^2} \frac{2}{m^2 (d-4)^2} \left(\frac{4 \pi \tilde{\mu}^2}{m^2} \right)^{2-d/2} \Gamma(3-d/2) \, {}_2F_1(1, 3-d/2; d/2-1;s/m^2) \, , \notag \\
    \label{eq:loopIntegrals}
}
where we used the integral representation of the beta function and the hypergeometric function in the Feynman parameter integrals:
\eqa{
    \int_0^1  x^a \bar{x}^b dx &=& \beta(a+1,b+1) \, , \\
    \int_0^1  x^a \bar{x}^b (1-zx)^c dx &=& \beta(a+1,b+1) \, {}_2F_1(-c,a+1;a+b+2;z) \, , \label{eq:BetaAndHyper}
}
with the notation $\bar{x} \equiv 1-x$.
We computed the $d$-dimensional expressions from scratch and checked the results up to and including $\mathcal{O}(\epsilon^0)$ with \texttt{Package-X} \cite{Patel:2015tea}.

\section{Details of the Phase Space Integration}
\label{app:PS}
The evaluation of $I_4$ goes as follows:
\eqa{I_4(a,b,c,d,e,f) &=& \int_0^1 dx dy \,  \frac{x^a \bar{x}^b y^c \bar{y}^d}{1-xy} {}_2F_1(1,e,f,x y) \notag \\
&=& \sum_{n=0}^\infty \frac{(1)_n (e)_n}{(f)_n} \frac{1}{n!} \int_0^1 dxdy \, \frac{x^{a+n} \bar{x}^b y^{c+n} \bar{y}^d}{1-xy} \notag \\
&=&  \sum_{n=0}^\infty \frac{(1)_n (e)_n}{(f)_n} \frac{\beta(c+n+1,d+1)}{n!} \int_0^1 dx \, x^{a+n} \bar{x}^b \,  {}_2F_1(1,c+n+1,c+d+n+2,x) \notag  \\
&=& \sum_{n=0}^\infty \frac{(1)_n (e)_n}{(f)_n} \frac{\beta(c+n+1,d+1) \beta(a+n+1,b+1)}{n!} \notag \\
& & \times{}_3F_2(1,c+n+1,a+n+1; c+d+n+2,a+b+n+2; 1) \, , \label{eq: 2F1Int}
}
where we used the series expansion of ${}_2F_1$
\eq{
{}_2F_1(a,b,c,z) = \sum_{n=0}^\infty \frac{(a)_n (b)_n}{(c)_n} \frac{z^n}{n!} \, ,
}
with the Pochammer symbol $(a)_n = \Gamma(a+n)/\Gamma(a)$.
For the parameter values that appear in our calculation, we find that the difference of the third and fifth argument of ${}_3F_2$ is either zero or a positive integer.
If we encounter a case where the difference is zero, we use
\eqa{
{}_3F_2(a, b, f; c, f; 1)  = {}_2F_1(a,b;c;1) = \frac{\Gamma(c) \Gamma(c - a - b)}{\Gamma(c-a) \Gamma(c-b)}
}
If the difference is a positive integer, we employ Clausen's series expansion \cite{Karlsson:1971}
\eq{
{}_3F_2(a, b, f+m; c, f; 1) = \frac{\Gamma(c)\Gamma(c-a-b)}{\Gamma(c-a)\Gamma(c-b)} \sum_{k=0}^m {m\choose k} \frac{(-1)^k (a)_k (b)_k}{(f)_k (1+a+b-c)_k}\, .
}
Plugging this back into~Eq.~\ref{eq: 2F1Int}, yields a sum that can be computed. The result is then expanded in~$\epsilon$ using \texttt{HypExp} \cite{Huber:2007dx}.


\section{Renormalization}
\label{app:renormalization}

In this appendix, we describe the details of the renormalization procedure, give the full list of evanescent operators necessary for our calculation, and give a collection of the relevant renormalization constants.

\subsection{Details of the Renormalization Procedure}
We consider the renormalized Lagrangian of the LEFT/WET, normalized as in the Bern basis:
\eqa{
\L_{\rm EFT} \equiv \L - \L_{\rm ren} = \frac{4G_F}{\sqrt{2}} \sum_{i,j} C_i Z^C_{ij}\, Z_{Q_j} Q_j \, ,
}
with the sum running over all relevant physical and evanescent operators. We define $\L_{\rm ren}$ here as containing all dimension-four or lower terms. While the renormalization constant $Z_{Q_k}$ accounts for the field-strength renormalization of the fields contained in the operators, the matrix $\hat{Z}^C$ describes the renormalization of the Wilson coefficients. Here, $Q_i$ and $C_i$ are renormalized operators and Wilson coefficients, respectively. In this article, we choose to regularize both UV and IR divergences using (naive) dimensional regularization, and to renormalize the Lagrangian in the $\overline{\text{MS}}$ scheme, with the Buras-Weisz prescription for the renormalization of evanescent operators~\cite{Buras:1989xd}.
As discussed in Sec.~\ref{sec:Calculation}, we obtain the contributions to the decay width by computing the imaginary part of double insertions of the corresponding Lagrangian terms. Schematically, we write
\eqa{
\Gamma \sim T\{ \L_{\rm EFT}(x) \L_{\rm EFT}(0) \} \, .
}
Writing this in terms of renormalized operators and Wilson coefficients yields
\eqa{
T\{\L_{\rm EFT}(x) \L_{\rm EFT}(0) \} \sim C_i C_j^\dagger Z^C_{ik} Z^C_{jl} Z_{Q_k} Z_{Q_l}\, T\{ Q_k(x) Q^\dagger_l(0) \} \, .
}
We then write
\eqa{
Z^C_{ij} = \delta_{ij} +  \delta Z^C_{ij} \, , \quad Z_{Q_K} = 1 + \delta Z_{Q_k} \, , 
}
where $\delta Z^C$ and $\delta Z_Q$ will give rise to the so-called counterterm contributions.
Note that in our calculation, we will only need to take into account $Z_{Q_k}$ when dealing with four-fermion operators. 
Since in the $\overline{\text{MS}}$-scheme the renormalization constants for the different quark flavors are all the same, we will use
\eqa{
Z_{Q_k} = \left(Z_q^{\overline{\text{MS}}}\right)^2 \, ,
}
with the usual $\overline{\text{MS}}$ field-strength renormalization constant $Z_q^{\overline{\text{MS}}}$. 
With this, we get
\begin{align}
T\{\L(x) \L(0) \}  \sim & \,  C_i  C_j^{\dagger} \, \left[ T\{ Q_i Q^\dagger_j \} + \delta Z^C_{ik} \, T\{ Q_k Q^\dagger_j \} + \delta Z^C_{jl} \, T\{ Q_i Q^\dagger_l \} + (\delta Z_{Q_i} + \delta Z_{Q_j})\, T\{ Q_i Q^\dagger_j \} \right ]\, \notag \\*
&+ \mathcal{O}(\tilde{\alpha}_s^2) \, , 
\label{eq:ren1}
\end{align}
where we used the fact that $\delta Z = \mathcal{O}(\tilde{\alpha}_s)$ in both types of counterterms. Let us remark that, thus far, this only takes into account the renormalization of higher-dimension operators and Wilson coefficients. However, when computing the imaginary parts of the double-insertion diagrams, one must also include all contributions from renormalization constants from within $\L_{\rm ren}$.

Let us now note that since we renormalized the Lagrangian in the $\overline{\text{MS}}$-scheme, the $b \to b$ matrix elements must include an LSZ factor of the ratio of the quark on-shell and $\overline{\text{MS}}$ renormalization constant $Z_b^{\text{OS}}/Z_b^{\overline{\text{MS}}}$. 
This ratio contains IR divergences necessary to cancel those of the four-particle cuts in which the gluon goes from the initial to the final $b$-quark.
Furthermore, when cutting through internal quark lines (of flavor $q$) one has to include the three-particle cuts with a $Z_q^{\text{OS}}/Z_q^{\overline{\text{MS}}}$ factor due to the fourth Cutkosky rule in Appendix~\ref{app:Cutkosky}. This contribution amounts then, up to $O(\alpha_s)$, to an additional counterterm of the form
\begin{equation}
\label{eq:ZqOS_Contribution}
\big(\delta Z_q^{\text{OS}} - \delta Z_q^{\overline{\text{MS}}}\big)
\vcenter{\hbox{
  \includegraphics[width=0.25\textwidth]{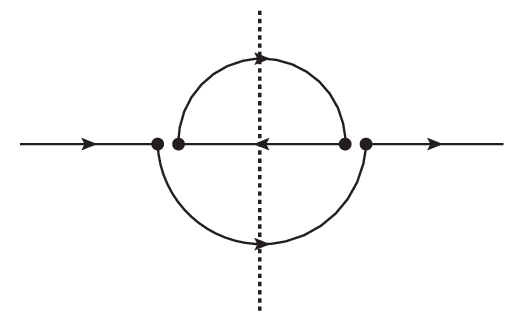}
}} \, .
\end{equation}
However, for a massless quark, the on-shell wave function renormalization constant is $Z_q^{\rm OS} = 1$, and thus $\delta Z_q^{\text{OS}}$ vanishes in~Eq.~(\ref{eq:ZqOS_Contribution}).

Let us now digress to connect this to the usual formalism based on computing individual $b$-decay amplitudes and then squaring them to obtain $\Gamma(b)$. 
In the cut-diagram approach, the insertion of two four-quark operators with flavor content $(\bar{q}_1b)(\bar{q}_2q_3)$ (with all $q_i$ being light) yields the following field-strength renormalization factors
\eqa{
\underbrace{\delta Z_b^{\text{OS}} - \delta Z_q^{\overline{\text{MS}}}}_{\text{LSZ}} + \underbrace{3 \delta Z_q^{\text{OS}} - 3 \delta Z_q^{\overline{\text{MS}}}}_{\text{SE 3p-cuts}} + \underbrace{4 \delta Z_q^{\overline{\text{MS}}}}_{\text{operator}} = \delta Z_b^{\text{OS}} +  3 \delta Z_q^{\text{OS}} = \delta Z_b^{\text{OS}}  \, ,
}
where the first terms are the standard LSZ factors, the second come from the extra factor in the fourth Cutkosky rule (i.e. three-particle cuts in SE-inserted diagrams), and the third term stems from $Z_{Q_i}Z_{Q_j}$ upon double insertion of $\L_{\rm EFT}$.
What in this approach is treated as cut internal lines, are external final states from the point of view of the amplitude. Hence, in the amplitude-level approach there there are no SE bubbles attached on the final-state quark lines, with which there is also no counterterm for the SE insertion. 
Much rather, these are absorbed into the standard LSZ factors, as explained in the derivation above. 
Thus, for the amplitude formalism, the counting of the field-strength renormalization constants after squaring the amplitude goes as follows:
\eqa{
\underbrace{\delta Z_b^{\text{OS}} + 3 \delta Z_q^{\text{OS}} - \delta 4 Z_q^{\overline{\text{MS}}}}_{\text{LSZ}} + \underbrace{4 \delta Z_q^{\overline{\text{MS}}}}_{\text{operator}} = \delta Z_b^{\text{OS}} +  3 \delta Z_q^{\text{OS}} = \delta Z_b^{\text{OS}}  \, .
}
Of course, the final result in both cases is the same, but it is nevertheless enlightening to see the connection between the two formalisms.

Finally, adding these factors to~Eq.~\ref{eq:ren1}, yields
\begin{align}
\text{Im} \bra{b} \mathcal{L}(x) \mathcal{L}(0) \ket{b} \sim &  
C_i C_j^{\dagger} \, \left[ \langle Q_i Q^\dagger_j \rangle + \delta Z^C_{ik} \, \langle Q_k Q^\dagger_j \rangle^{\text{tree}} + \delta Z^C_{jl} \, \langle Q_i Q^\dagger_l \rangle^{\text{tree}} + \delta Z_b^{\text{OS}} \, \langle Q_i Q^\dagger_j \rangle^{\text{tree}} \right ]\, \notag \\*
&+ \mathcal{O}(\tilde{\alpha}_s^2) \, .
\end{align}
Furthermore, we can split up the counterterm contributions in between those that cancel the divergences coming from the $CC$ insertions and the ones coming from the penguin insertions. To this end we write:
\eqa{
\delta \hat{Z}^{C} = \delta \hat{Z}_{CC} + \delta \hat{Z}_P \, ,
}
where  $\delta \hat{Z}_{CC}$ together with all contributions from the wavefunction renormalization $\delta Z^{\text{OS}}_b$ will take care of the UV divergences (and partly also the IR divergences) coming from the $CC$ insertions, while $\delta \hat{Z}_P$ will render the penguin-like contributions finite.
Note that at NLO, concerning the counterterms from $\L_{\rm ren}$, we only need the aforementioned quark field-strength renormalization constants.
For completeness we state them here in the $\overline{\text{MS}}$ and on-shell scheme:
\eqa{
Z_q^{\overline{\text{MS}}} &=& 1 - C_F \, \frac{\tilde{\alpha}_s}{\epsilon} \, ,\\
Z_b^{\text{OS}} &=& 1 - C_F \, \tilde{\alpha}_s \left(\frac{3}{\epsilon} + 6 \log \frac{\mu}{m_b} + 4 \right) \, .
}
The difference in the $1/\epsilon$ factors between the two schemes is explained by the fact that $Z_b^{\text{OS}}$ includes also IR divergences.

\subsection{Evanescent Operators}
Here, we collect the full set of evanescent operators necessary for our calculation separated by the decay channel where they appear. Additionally, we split the evanescent operators into Chisholm-evanescent, Fierz-evanescent, or Chisholm-Fierz-evanescent operators depending on whether they are necessary to generalize the four-dimensional Chisholm, Fierz, or a combination of both identities.

The usual prescription for defining Chisholm-evanescent operators consists in applying some iteration of the Greek projections \cite{Tracas:1982gp,Buras:2000if} as presented in detail, for instance, in Ref.~\cite{Dekens:2019ept}. However, the straightforward application of these methods will yield certain combinations of the usual scalar, vector, and tensor chiral structures which would be troublesome for our calculation. This is because we need to include counterterm diagrams with insertions of evanescent operators, and thus would end up having to compute Dirac traces of these chiral structures in $d$ dimensions, incurring in the issue of ill-defined traces of $\gamma_5$. In order to avoid this problem, we instead define the Chisholm-evanescent operators as preserving the Bern-like structure (i.e. avoiding all $\gamma_5$ in at least one of the currents), based instead on an irreducible basis of higher Dirac structures and physical operators. Concerning the $\mathcal{O}(\epsilon)$ pieces of the evanescent operators, for which there is a freedom usually associated with the choice of one or another \textit{evanescent scheme}, we fix them by explicitly demanding that the resulting evanescent basis be a linear combination (with no additional physical-operator contributions) of the common Greek-projection evanescent basis as defined in, e.g. Refs.~\cite{Buras:2000if,Meiser:2024zea, Morell:2024aml}, or equivalently to the choice of $a_{ev} = b_{ev} = \dots = 1$ in Ref.~\cite{Dekens:2019ept}. This ensures that the evanescent scheme defined by the two bases, i.e. the finite terms due to evanescent insertions, will be the same~\cite{Dugan:1990df,Herrlich:1994kh,Gorbahn:2004my}.
Equivalently, this can also be achieved through the simplification of all higher Dirac structures with the direct use of the Greek-projection evanescents, and later rewriting into the Bern-like structure.

\subsubsection{Case 1 ($b \to d\bar{u}u$) } 
For the $b \to d \bar{u} u$ channel, we only need Chisholm-evanescent operators. Following the prescription outlined in the text above, we find the following definitions:
\eqa{
E_{C;5}^{S, \, dbuu} &=& \left[\bar{d} \gamma_{\mu_1 \mu_2 \mu_3 \mu_4 \mu_5} P_L b \right] \left[\bar{u} \gamma^{\mu_1 \mu_2 \mu_3 \mu_4 \mu_5} u \right] + (64 - 112 \epsilon) \mathcal{O}_1^{dbuu} + (-20 + 16 \epsilon) \mathcal{O}_3^{dbuu} \, , \\
E_{C;5}^{O, \, dbuu} &=& \left[\bar{d} \gamma_{\mu_1 \mu_2 \mu_3 \mu_4 \mu_5} P_L T^A b \right] \left[\bar{u} \gamma^{\mu_1 \mu_2 \mu_3 \mu_4 \mu_5} T^A u \right] + (64 - 112 \epsilon) \mathcal{O}_2^{dbuu} + (-20 + 16 \epsilon) \mathcal{O}_4^{dbuu} \, , \\
E_{C;6}^{S, \, dbuu} &=&  \left[\bar{d} \gamma_{\mu_1 \mu_2 \mu_3 \mu_4 \mu_5 \mu_6} P_R b \right] \left[\bar{u} \gamma^{\mu_1 \mu_2 \mu_3 \mu_4 \mu_5 \mu_6} u \right] + (256 - 1280 \epsilon) \mathcal{O}_5^{dbuu} + (-64 +384 \epsilon) \mathcal{O}_7^{dbuu} \notag \\
& & + (-20 + 36 \epsilon) \mathcal{O}_9^{dbuu} \, , \\
E_{C;6}^{O, \, dbuu} &=&  \left[\bar{d} \gamma_{\mu_1 \mu_2 \mu_3 \mu_4 \mu_5 \mu_6} P_R T^A b \right] \left[\bar{u} \gamma^{\mu_1 \mu_2 \mu_3 \mu_4 \mu_5 \mu_6} T^A u \right] + (256 - 1280 \epsilon) \mathcal{O}_6^{dbuu} + (-64 +384 \epsilon) \mathcal{O}_8^{dbuu} \notag \\
& & + (-20 + 36 \epsilon) \mathcal{O}_{10}^{dbuu} \, ,
}
where the subscript $C$ indicates that these are Chisholm-evanescent operators. The chiral-opposite operators can be easily obtained by $P_L \leftrightarrow P_R$.

\subsubsection{Case 2 and 3: $b \to d \bar{s}d$ and $b \to d \bar{d}d$} 
The $b \to d\bar{s}d$ channel contributes only with $CC$-insertions, which yield Chisholm-evanescent structures analogous to those of $b \to d \bar{u}u$ $CC$ insertions. However, and differing from the latter, color-octet bilinear structures are Fierz-redundant in the $b \to d\bar{s}d$ case. Given that these structures are still produced in intermediate states of the calculation, we need to reduce them to color-singlet structures using Fierz-evanescent operators. This applies to the Chisholm-evanescent structures all the same, thus changing the definition of the corresponding operators. The evanescent operators that only generalize Fierz identities carry the subscript $F$, while the combined Chisholm-Fierz-evanescent operators carry the subscript $CF$.
\eqa{
E_{F;2}^{dbds} &=&
\mathcal{O}_2^{dbds} - \mathcal{O}_1^{dbds} \left(-\frac{5 \epsilon }{18}-\frac{1}{3} \right)-\mathcal{O}_3^{dbds} \left(\frac{\epsilon }{36}+\frac{1}{24}\right)-\mathcal{O}_5^{dbds} \left(-\frac{14 \epsilon }{9}-\frac{4}{3}\right) -\mathcal{O}_7^{dbds} \left(\frac{13 \epsilon }{18}+\frac{1}{3}\right)
\notag \\*
& & -
\mathcal{O}_9^{dbds} \left(\frac{\epsilon }{18}+\frac{1}{48}\right) \, , \\
E_{F;4}^{dbds} &=& \mathcal{O}_4^{dbds} + \mathcal{O}_1^{dbds} \left(\frac{34 \epsilon }{9}+\frac{8}{3}\right)+\frac{1}{18} \mathcal{O}_3^{dbds} (-5 \epsilon -9)+\frac{8}{9} \mathcal{O}_5^{dbds} (13 \epsilon +6)-\frac{2}{9} \mathcal{O}_7^{dbds} (19 \epsilon +6) \notag \\*
& & +\frac{1}{36} \mathcal{O}_9^{dbds} (-11 \epsilon -3) \, , \\
E_{F;6}^{dbds} &=& \mathcal{O}_6^{dbds} + \mathcal{O}_1^{dbds} \left(\frac{5 \epsilon }{36}+\frac{1}{3}\right)+\frac{1}{144} \mathcal{O}_3^{dbds} (-2 \epsilon -3)+\frac{1}{36} \mathcal{O}_5^{dbds} (4 \epsilon +3)+\mathcal{O}_7^{dbds} \left(\frac{\epsilon }{18}+\frac{7}{48}\right) \notag \\*
& & +\mathcal{O}_9^{dbds} \frac{7 \epsilon +6}{1152} \, , \\
E_{F;8}^{dbds} &=& \mathcal{O}_8^{dbds}-2 \mathcal{O}_1^{dbds} \epsilon +\frac{\mathcal{O}_3^{dbds} \epsilon }{8}+\mathcal{O}_5^{dbds} \left(-\frac{20 \epsilon }{3}-1\right)+\mathcal{O}_7^{dbds} \left(\frac{8 \epsilon }{3}+\frac{11}{12}\right)+ \mathcal{O}_9^{dbds}  \frac{19 \epsilon +6}{96} \, , \notag\\
\\
E_{F;10}^{dbds} &=& \mathcal{O}_{10}^{dbds} + \frac{4}{9} \mathcal{O}_1^{dbds} (101 \epsilon +12)+\frac{1}{9} \mathcal{O}_3^{dbds} (-26 \epsilon -3)+\frac{32}{9} \mathcal{O}_5^{dbds} (32 \epsilon +3)-\frac{8}{9} \mathcal{O}_7^{dbds} (53 \epsilon +3) \notag \\*
& & +\mathcal{O}_9^{dbds} \left(-\frac{25 \epsilon }{9}-\frac{1}{2}\right) \, , \\
E_{CF;5}^{dbds} &=& \left[\bar{d}\gamma_{\mu_1 \mu_2 \mu_3 \mu_4 \mu_5}P_L T^A b \right]\left[\bar{d} \gamma^{\mu_1 \mu_2 \mu_3 \mu_4 \mu_5} T^A s \right]+ \mathcal{O}_1^{dbds} \left(\frac{472 \epsilon }{9}+32\right)-\frac{2}{9} \mathcal{O}_3^{dbds} (2 \epsilon +33)\notag \\*
& & +\frac{32}{9} \mathcal{O}_5^{dbds} (55 \epsilon +6) -\frac{8}{9} \mathcal{O}_7^{dbds} (61 \epsilon +6)+\frac{1}{9} \mathcal{O}_9^{dbds} (-32 \epsilon -3) \, , \\
E_{CF;6}^{dbds} &=& \left[\bar{d}\gamma_{\mu_1 \mu_2 \mu_3 \mu_4 \mu_5 \mu_6}P_L T^A b \right]\left[\bar{d}\gamma^{\mu_1 \mu_2 \mu_3 \mu_4 \mu_5 \mu_6} T^A s \right] + \frac{16}{9} \mathcal{O}_1^{dbds} (545 \epsilon +12)-\frac{4}{9} \mathcal{O}_3^{dbds} (137 \epsilon +3) \notag \\*
& & +\frac{64}{9} \mathcal{O}_5^{dbds} (271 \epsilon +18)+\mathcal{O}_7^{dbds} \left(-\frac{7696 \epsilon }{9}-32\right)-\frac{2}{9} \mathcal{O}_9^{dbds} (197 \epsilon +33) \, .
} 
For the $b \to d\bar{d}d$ channel, the evanescent operators are obtained from the previous $b \to d\bar{s}d$ definitions by swapping $s \to d$.

\subsection{Collection of Renormalization Constants}
Here, we collect all relevant one-loop renormalization constants and explain the intricacies of the renormalization of some of the contributions.

\subsubsection{Case 1: $b \to u \bar{u} d$} 
\paragraph{$CC$ Insertions}
The $CC$ insertions are rather straightforward to renormalize. To begin with, their renormalization is flavor-diagonal in the sense that the insertion an operator with a fixed flavor structure in one side of a cut diagram must go with an identical flavor structure (up to Fierz transformations) on the other side. Since, for definiteness, we focus on $b \to d \bar{u} u$, we only present the $CC$ renormalization constants for operators with $dbuu$ flavor structure:
\eqa{
\delta \hat{Z}^{CC} = 
    \left(
    \begin{array}{cc}
    \delta \hat{Z}^{CC}_{1-4} & 0_{4 \times 8} \\
    0_{6 \times 6} & \delta \hat{Z}^{CC}_{5-10}
    \end{array} 
    \right) \, ,
}
with the subblocks 
\eqa{
\delta \hat{Z}^{CC}_{1-4} &=& \frac{\tilde{\alpha}_s}{\epsilon} \left(
\begin{array}{cccccc}
 0 & -10 & 0 & 1 & 0 & 0 \\
 -\frac{20}{9} & -\frac{26}{3} & \frac{2}{9} & \frac{5}{12} & 0 & 0 \\
 0 & -64 & 0 & 10 & 0 & 1 \\
 -\frac{128}{9} & -\frac{80}{3} & \frac{20}{9} & -\frac{1}{3} & \frac{2}{9} & \frac{5}{12} \\
\end{array}
\right) \, , \\[8pt]
\delta \hat{Z}^{CC}_{5-10} &=& \frac{\tilde{\alpha}_s}{\epsilon} \left(
\begin{array}{cccccccc}
 -8 & 0 & 0 & -1 & 0 & 0 & 0 & 0 \\
 0 & 1 & -\frac{2}{9} & -\frac{5}{12} & 0 & 0 & 0 & 0 \\
 0 & 16 & \frac{8}{3} & -16 & 0 & -1 & 0 & 0 \\
 \frac{32}{9} & \frac{20}{3} & -\frac{32}{9} & -13 & -\frac{2}{9} & -\frac{5}{12} & 0 & 0 \\
 0 & -256 & -\frac{512}{3} & 192 & -8 & 16 & 0 & 1 \\
 -\frac{512}{9} & -\frac{320}{3} & \frac{128}{3} & \frac{592}{3} & \frac{32}{9} & \frac{23}{3} & \frac{2}{9} & \frac{5}{12} \\
\end{array}
\right) \, ,
}
with the following ordering of the operators:
\eqa{
\vec{\mathcal{O}}_{1-4} &=& \left(\mathcal{O}_1^{dbuu}, \mathcal{O}_2^{dbuu}, \mathcal{O}_3^{dbuu}, \mathcal{O}_4^{dbuu}, E_5^{S, \, dbuu}, E_5^{O, \, dbuu} \right) \, , \\
\vec{\mathcal{O}}_{5-10} &=& \left(\mathcal{O}_5^{dbuu}, \mathcal{O}_6^{dbuu}, \mathcal{O}_7^{dbuu}, \mathcal{O}_8^{dbuu}, \mathcal{O}_9^{dbuu}, \mathcal{O}_{10}^{dbuu}, E_6^{S, \, dbuu}, E_6^{O, \, dbuu} \right) \, .
}
Note that, in these matrices, the rows only run over the physical operators of the basis, while the columns also run over the evanescent operators. The physical part of the counter terms corresponds to the contribution to the one-loop ADM coming from CC insertions, and we indeed reproduce the class III one-loop ADM in Ref.~\cite{Aebischer:2017gaw}.

\paragraph{Penguin Insertions}
The PP insertions are slightly more involved in their renormalization. Due to the structure of our operators, we only need to consider closed-penguin contributions for $q = u, s, c$, while open-penguin insertions are also necessary for $q = d, b$, and turn out to be the only relevant contributions due to the color structure of the operators in the basis. 
The full penguin-diagram renormalization matrix reads: 
\eqa{
\delta \hat{Z}_{P} = 
\left(\begin{array}{cccccc}
    \delta \hat{Z}_P^{uu} & \delta \hat{Z}_P^{us} & \delta \hat{Z}_P^{uc} & \delta \hat{Z}_P^{ud} & \delta \hat{Z}_P^{ub} & 0   \\
    \delta \hat{Z}_P^{su} & \delta \hat{Z}_P^{ss} & \delta \hat{Z}_P^{sc} & \delta \hat{Z}_P^{sd} & \delta \hat{Z}_P^{sb} & 0    \\
    \delta \hat{Z}_P^{cu} & \delta \hat{Z}_P^{cs} & \delta \hat{Z}_P^{cc} & \delta \hat{Z}_P^{cd} & \delta \hat{Z}_P^{cb} & \delta \hat{Z}_P^{cD} \\
    \delta \hat{Z}_P^{du} & \delta \hat{Z}_P^{ds} & \delta \hat{Z}_P^{dc} & \delta \hat{Z}_P^{dd} & \delta \hat{Z}_P^{db} & 0   \\
    \delta \hat{Z}_P^{bu} & \delta \hat{Z}_P^{bs} & \delta \hat{Z}_P^{bc} & \delta \hat{Z}_P^{bd} & \delta \hat{Z}_P^{bb} & \delta \hat{Z}_P^{bD}
\end{array}
\right) \, . \label{eq:renPeng}
}
The superscripts in these matrices are such that $\delta \hat{Z}_P^{\psi_1\psi_2}$ stems from a one-loop penguin diagram with an $\psi_1$-quark loop, and projects onto the structure of a $db\psi_2\psi_2$ operator.
Note that for penguins with massive quarks we also need counterterms with the structure of dipole operators, $\delta \hat{Z}_P^{\psi_1D}$. 

For the $b \to du \bar{u}$ channel, out of all the renormalization matrix, we are only interested in the sub-blocks that project onto $dbuu$ structures, as they are the only ones that can be featured in counterterm diagrams with a $du\bar{u}$ cut: 
\eqa{
\delta \hat{Z}_P^{uu} = \delta \hat{Z}_P^{su} = \delta \hat{Z}_P^{cu} = \frac{\tilde{\alpha
}_s}{\epsilon} \left(
\begin{array}{ccc}
 0 & 0 & 0_{1 \times 8} \\
 0 & \frac{2}{3} & 0_{1 \times 8} \\
 0 & 0  & 0_{1 \times 8} \\
 0 & \frac{20}{3}  & 0_{1 \times 8} \\
 0_{6 \times 1} & 0_{6 \times 1} &  0_{6 \times 8}
\end{array}
\right) \, , \quad
\delta\hat{Z}_P^{du} = \delta\hat{Z}_P^{bu} = \frac{\tilde{\alpha}_s}{\epsilon} \left(
\begin{array}{cccccccccc}
 0 & \frac{2}{3} & 0_{8 \times 1}  \\
 0 & \frac{32}{3} & 0_{8 \times 1} \\
 0 & -\frac{1}{3} & 0_{8 \times 1} \\
 0 & 0 & 0_{8 \times 1} \\
 0 & -\frac{16}{3} & 0_{8 \times 1} \\
\end{array}
\right) \, .
}
The additional dipole-operator counterterms required in diagrams with massive loops are:
\eqa{
\delta \hat{Z}^{cD} = \frac{\tilde{\alpha}_s}{\epsilon} \left(
\begin{array}{c}
 0_{7 \times 1} \\
 -2 z_c \\
 0 \\
 32 z_c \\
\end{array}
\right) \, , \quad 
\delta \hat{Z}^{bD} = \frac{\tilde{\alpha}_s} {\epsilon}\left(
\begin{array}{c}
 0_{2 \times 1} \\
 \frac{1}{2} \\
 -2 \\
 64 \\
\end{array}
\right) \, .
}
These renormalization constants, together with the $CC$ renormalization constants, reproduce the corresponding parts of the one-loop ADMs of the Class V operators in Ref.~\cite{Aebischer:2017gaw}.

\subsubsection{Case 2: $b \to d \bar{s} d$} 
For $b \to d\bar{s}d$, what we refer to as Case 2 in the main text, there are only $CC$ insertions. The renormalization constants read:
\eqa{
\delta\hat{Z}^{dbds} = \frac{\tilde{\alpha}_s}{\epsilon}
\left(
\begin{array}{cccccccccccc}
 \frac{2}{3} & \frac{1}{12} & 8 & -2 & -\frac{1}{8} & -10 & 1 & 0 & 0 & 0 & 0 & 0 \\
 -\frac{16}{3} & \frac{7}{3} & 32 & -8 & -\frac{1}{2} & 0 & -10 & 0 & 0 & 0 & 1 & 0 \\
 0 & 0 & -9 & \frac{11}{12} & \frac{1}{16} & 0 & 0 & 0 & -1 & 0 & 0 & 0 \\
 0 & 0 & -\frac{20}{3} & \frac{37}{3} & \frac{5}{12} & 0 & 0 & 16 & -16 & -1 & 0 & 0 \\
 0 & 0 & \frac{128}{3} & -\frac{800}{3} & -\frac{32}{3} & 0 & 0 & 0 & 128 & -4 & 0 & 1 \\
\end{array}
\right) \, ,
}
with the ordering
\eqa{
\vec{\mathcal{O}}^{dbds} = \left(\mathcal{O}_1^{dbds}, \mathcal{O}_3^{dbds}, \mathcal{O}_5^{dbds}, \mathcal{O}_7^{dbds}, \mathcal{O}_9^{dbds}, E_{F;2}^{dbds}, E_{F;4}^{dbds}, E_{F;6}^{dbds}, E_{F;8}^{dbds}, E_{F;10}^{dbds}, E_{CF;5}^{dbds}, E_{CF;6}^{dbds} \right) \, .
\quad
}
Again, the rows only run over the 5 physical operators in the basis, while the columns also run over the relevant evanescent operators. The physical part reproduces the Class IV one-loop QCD ADM of Ref.~\cite{Aebischer:2017gaw}.

\subsubsection{Case 3: $b \to d\bar{d}d$}
The renormalization of the $CC$ insertions in Case 3 works exactly in the same way as for Case 2. However, they also require the renormalization constants for the penguin insertions. For this channel, we need the sub-blocks of the matrix in~Eq.~\ref{eq:renPeng} that project onto $dbbb$ structures:
\eqa{
\delta \hat{Z}_P^{ud} = \delta \hat{Z}_P^{sd} = \delta \hat{Z}_P^{cd} &=& \frac{\tilde{\alpha}_s}{\epsilon}\left(
\begin{array}{cccccc}
 0 & 0 & 0 & 0 & 0 & 0 \\
 -\frac{2}{9} & \frac{1}{36} & -\frac{8}{9} & \frac{2}{9} & \frac{1}{72} & \frac{2}{3} \\
 0 & 0 & 0 & 0 & 0 & 0 \\
 -\frac{20}{9} & \frac{5}{18} & -\frac{80}{9} & \frac{20}{9} & \frac{5}{36} & \frac{20}{3} \\
 0_{6 \times 1} & 0_{6 \times 1} & 0_{6 \times 1} & 0_{6 \times 1} & 0_{6 \times 1} & 0_{6 \times 1} 
\end{array}
\right)\, , \\[8pt]
\delta \hat{Z}_P^{dd} = \delta \hat{Z}_P^{bd} &=& \frac{\tilde{\alpha}_s}{\epsilon}\left(
\begin{array}{cccccc}
 -\frac{2}{9} & \frac{1}{36} & -\frac{8}{9} & \frac{2}{9} & \frac{1}{72} & \frac{2}{3} \\
 -\frac{32}{9} & \frac{4}{9} & -\frac{128}{9} & \frac{32}{9} & \frac{2}{9} & \frac{32}{3} \\
 \frac{1}{9} & -\frac{1}{72} & \frac{4}{9} & -\frac{1}{9} & -\frac{1}{144} & -\frac{1}{3} \\
 0 & 0 & 0 & 0 & 0 & 0 \\
 \frac{16}{9} & -\frac{2}{9} & \frac{64}{9} & -\frac{16}{9} & -\frac{1}{9} & -\frac{16}{3} \\
\end{array}
\right) \, .
}
The ordering of the columns is
\eqa{
\vec{\mathcal{O}}^{dbdd} = \left(\mathcal{O}_1^{dbdd}, \mathcal{O}_3^{dbdd}, \mathcal{O}_5^{dbdd}, \mathcal{O}_7^{dbdd}, \mathcal{O}_9^{dbdd}, E_{F;2}^{dbdd} \right) \, ,
}
and the rows run only over the physical operators. The physical part of these counterterm matrices reproduces the corresponding parts of the one-loop ADM in of the class V operators in Ref.~\cite{Aebischer:2017gaw}.

\section{Leading Order Result for $b \to d\bar{c}c$}
\label{app:btodcc}
The LO result for the channel $b \to d \bar{c}c$ can be conveniently expressed with the following two functions
\eqa{
f_{cc}(z_c) &=& \Delta_c \left(1 - 14 z_c^2 - 2 z_c^4 - 12 z_c^6 \right)+24 \Lambda_c \left(1 - z_c^4\right) z_c^4 \, , \\
g_{cc}(z_c) &=& 2 z_c \left[\Delta_c \left(-1 - 5 z_c^2 + 6 z_c^4 \right) + 6 z_c^2 \left(2 z_c^4-2 z_c^2+1\right) \Lambda_c  \right] \, .
}
The rest of the entries can be written in terms of linear combinations of these two:
\eqa{
h_{cc;1}(z_c) &=& f_{cc}(z_c) - 2 \, z_c \, g_{cc}(z_c) \, , \\
h_{cc;2}(z_c) &=& f_{cc}(z_c) + 4 \, z_c \, g_{cc}(z_c) \, , \\
h_{cc;3}(z_c) &=& 65 \, f_{cc}(z_c) + 32 \, z_c \, g_{cc}(z_c) \, \\
h_{cc;4}(z_c) &=& 17 \, f_{cc}(z_c) - 16 \, z_c \, g_{cc}(z_c) \, .
}
Using these, the LO result reads
\eqa{
& & \hat{G}_{CC}^{(0)}[d\bar{c}c] = \notag \\*[6pt]
& &  \left(
\begin{array}{cccccccccc}
 6h_{cc;1} & 0 & 60h_{cc;1} & 0 & 0 & 0 & -36g_{cc} & 0 & 576g_{cc} & 0 \\
 0 & \frac{4}{3} h_{cc;1} & 0 & \frac{40}{3} h_{cc;1} & 0 & 0 & 0 & -8g_{cc} & 0 & 128g_{cc} \\
 60h_{cc;1} & 0 & 48 h_{cc;4} & 0 & 0 & 0 & -360g_{cc} & 0 & 6624g_{cc} & 0 \\
 0 & \frac{40}{3} h_{cc;1} & 0 & \frac{32}{3} h_{cc;4} & 0 & 0 & 0 & -80g_{cc} & 0 & 1472g_{cc} \\
 0 & 0 & 0 & 0 & \frac{3}{2} h_{cc;2} & 0 & 0 & 0 & 60 h_{cc;2} & 0 \\
 0 & 0 & 0 & 0 & 0 & \frac{1}{3} h_{cc;2} & 0 & 0 & 0 & \frac{40}{3} h_{cc;2} \\
 -36g_{cc} & 0 & -360g_{cc} & 0 & 0 & 0 & 36 f_{cc} & 0 & -576 f_{cc} & 0 \\
 0 & -8g_{cc} & 0 & -80g_{cc} & 0 & 0 & 0 & 8 f_{cc} & 0 & -128 f_{cc} \\
 576g_{cc} & 0 & 6624g_{cc} & 0 & 60 h_{cc;2} & 0 & -576 f_{cc} & 0 & 192 h_{cc;3} & 0 \\
 0 & 128g_{cc} & 0 & 1472g_{cc} & 0 & \frac{40}{3} h_{cc;2} & 0 & -128 f_{cc} & 0 & \frac{128}{3}  h_{cc;3} \\
\end{array}
\right) \notag \, , \\*
}
where we suppressed the argument of the functions in all the entries for readability.
We have compared this result to Ref.~\cite{Lang:2025ios} and found agreement. 

\bibliographystyle{JHEP}
\bibliography{references}

\end{document}